\documentclass[letterpaper,twocolumn,10pt]{article}
\usepackage{usenix2019_v3}
  
\usepackage{listings}
\usepackage{fancyvrb}
\usepackage{graphicx}
\usepackage[all]{hypcap}
\usepackage{mathptmx}
\usepackage[scaled]{couriers}
\usepackage{balance}
\usepackage{multirow}
\usepackage{epsfig}
\usepackage[linesnumbered,ruled,vlined]{algorithm2e}

\usepackage{blindtext}

\begin{document}

\date{}

\title{\Large \bf BigBen: Telemetry Processing for Internet-wide Event Monitoring}
\author{
{\rm Meenakshi Syamkumar}\\
University of Wisconsin-Madison
\and
{\rm Yugali Gullapalli}\\
University of Wisconsin-Madison
\and
{\rm Wei Tang}\\
University of Wisconsin-Madison
\and
{\rm Paul Barford}\\
University of Wisconsin-Madison
\and
{\rm Joel Sommers}\\
Colgate University
} 

\maketitle

\begin{abstract}This paper describes {\em BigBen}, a network telemetry processing system designed to enable accurate and timely reporting of Internet events ({\em e.g.,} outages, attacks and configuration changes).  BigBen is distinct from other Internet-wide event detection systems in its use of passive measurements of Network Time Protocol (NTP) traffic.  We describe the architecture of BigBen, which includes {\em (i)} a distributed NTP traffic collection component, {\em (ii)} an Extract Transform Load (ETL) component, {\em (iii)} an event identification component, and {\em (iv)} a visualization and reporting component.  We also describe a cloud-based implementation of BigBen developed to process large NTP data sets and provide daily event reporting.  We demonstrate BigBen on a 15.5TB corpus of NTP data.  We show that our implementation is efficient and could support hourly event reporting.  We show that BigBen identifies a wide range of Internet events characterized by their location, scope and duration.  We compare the events detected by BigBen vs. events detected by a large active probe-based detection system.  We find only modest overlap and show how BigBen provides details on events that are not available from active measurements. Finally, we report on the perspective that BigBen provides on Internet events that were reported by third parties.  In each case, BigBen confirms the event and provides details that were not available in prior reports, highlighting the utility of the passive, NTP-based approach.

\end{abstract}

\vspace{-0.3cm}
\section{Introduction}\label{sec:intro}\vspace{-0.2cm}
The dynamic nature of the Internet has been well documented over the years.  Events such as network reconfigurations, flash crowds, outages, persistent congestion and attacks of various sorts are commonplace and have been shown to have diverse spatio-temporal characteristics.  These events are of on-going interest from a research perspective, and have significant implications for day-to-day network management and operations.

The fact that events can occur at any time and in any place in the Internet complicates the task of {\em detection}. The {\em service provider perspective} on event detection focuses on a single infrastructure and assumes unfettered access to diverse measurement data such as active probe-based measurements, SNMP-based measurements, packet traces, flow-export data, and logs from devices deployed in the network.  In contrast, the {\em Internet-wide perspective} on event detection assumes that data must be collected outside of (most) network perimeters, typically using active probe-based methods.  Beyond gathering measurement data, challenges in event detection from both perspectives include {\em (i)} managing, fusing and processing potentially very large amounts of data, {\em (ii)} identifying events accurately and in a timely fashion and, {\em (iii)} organizing, refining and visualizing details of identified events.  Prior studies on event detection have focused on a variety of these issues and in particular on measurements and identification methods as described in Section~\ref{sec:relwork}.

In this paper, we describe a network telemetry processing and reporting system that we call {\em BigBen}, which is designed to provide an Internet-wide perspective on events.  The high-level goal of BigBen is an efficient processing platform that supports {\em timely} and {\em accurate} event reporting based on measurements collected {\em passively}.   The utility of event detection based on passive measurements is that it offers the opportunity to obviate standard challenges of active probe-based methods, which include management overhead, traffic overhead, potential for blocked probes and limited information provided by probes.  However, this approach raises the question of what passive measurement data might be available that can provide useful insights into internet-wide events?

To achieve our high-level goals we build on prior work that identifies the Network Time Protocol (NTP) as a compelling source of Internet-wide measurement data~\cite{Durairajan2015HotNets}.  In particular, NTP is the only on-by-default protocol in the Internet, which means that measurements of client connections across the globe can be passively collected at NTP servers.  Further, the timing-based nature of NTP data provides a means for extracting one way delays (OWDs) between clients and servers~\cite{Timeweaver18}, which enables events to be identified {\em e.g.,} by looking for changes in OWDs~\cite{Tezzeract18}.  

Our architecture for BigBen is designed for Internet-wide event detection based on NTP and includes {\em (i)} a distributed NTP traffic collection component, {\em (ii)} an Extract Transform Load (ETL) component, {\em (iii)} an event identification component, and {\em (iv)} a visualization and reporting component.   The design and interfaces between each component are modular and validate the importance of scalability, data integrity, extensibility and manageability in the system.  Details of the architecture for BigBen are provided in Section~\ref{sec:methods}.

We demonstrate the efficacy and utility of our design by developing a prototype implementation of BigBen.  To support timely ({\em e.g.,} daily or hourly) event reporting, a key requirement of the implementation is the ability to manage and process hundreds of GB to TB of NTP data per day~\cite{Durairajan2015HotNets}.  Handling data at this scale is a significant engineering challenge and in this sense BigBen is related to network telemetry systems such as~\cite{Sonata18}.   Our implementation utilizes a file-based approach for data management and Apache Spark for processing.  Our implementation also includes an event detector based on Robust Principal Components Analysis, which was shown to be effective for detecting spikes in NTP data in~\cite{Tezzeract18}.  However, recognizing the challenges in assessing event detection accuracy (due to the significant lack of Internet-wide ground truth~\footnote{Historically, reporting of events by ISPs, affected entities or third parties is rare, which is important motivation for our work.}), we emphasize that other detectors can easily be substituted into the system or run in parallel with the current detector.  Details of BigBen's implementation are provided in Section~\ref{sec:methods}.

We demonstrate BigBen through a study of Internet events based on data collected over a 5 month period from January through May, 2019.  For this study, we collected data on a daily basis from 16 NTP servers in 7 locations in the US resulting 15.5TB of total data.  Our implementations of the ETL, event identification and reporting components of BigBen were deployed in CloudLab~\cite{Cloudlab}.   Our system ran on a single (multi-core) node and was able to process data contributed by all 16 servers over a 24-hour (about 100 GB) in about 1 hour and 15 minutes.  This easily accomplished our design goal of being able to process gigabytes of raw NTP data and extract Internet events on a daily basis.  Moreover, the system can easily be extended to accommodate contributions from additional servers.  

Next, we drill down on events that are identified in the data.  We categorize events based on the fidelity of the signals in the NTP data.  For IPv4, we find that about 100K events per day are detected and that these events are indicated in about 60K prefixes.  For IPv6, we find that about 120 events per day are detected and that these events are indicated in about 90 prefixes.  We find that IPv4 events range in duration from 32 seconds to 27 hours, with an average event duration of 2.7 hours. We find that IPv6 events range in duration from 64 seconds to 24 hours, with an average event duration of 3.5 hours.   To the best of our knowledge, this is the first report of network events in IPv6 address space.  It is important to note that the fidelity of these measurements on the order of seconds is much more precise than what is possible from active probe-based systems ({\em e.g.,}~\cite{Heidemann08a_200802,quan2013trinocular}) that periodically scan the entire Internet.

To provide perspective on the events identified by BigBen,  we compare with reports of events identified by ISI's Trinocular system~\cite{Heidemann08a_200802,quan2013trinocular} that took place during the same time as our data collection.  While Trinocular covers substantially more /24 prefixes (4.2M vs. 488K), 76K prefixes are visible in the NTP data that are not visible in the ISI data.  Overall, Trinocular reports about 120K events per day vs. the 100K events per day reported by BigBen when aggregated over all signal levels.  We find relatively low event match rates between the two data sets, which we attribute to the relatively low overlap on /24 prefixes (only about 40K).  We argue that these results highlight the utility and complementary nature of BigBen vs. active probe-based systems for Internet-wide event detection.

Finally, we investigate how BigBen views events that have been reported by third parties during our data collection period.  We posit that these reports provide a measure of validation for the implemented event detector.  In each case, we find that BigBen detects an event that corresponds with what was reported.  We highlight how the events manifest in our data, and show that BigBen often provides an expanded perspective over what was reported.

In summary, the two main contributions of this paper are as follows.  First, we describe the architecture and implementation of BigBen, a system for Internet-wide event detection based on passive collection of NTP measurements.  Second, we demonstrate the capabilities of BigBen on a large corpus of NTP measurements collected over a 5 month period.  Our results show that BigBen is capable of detecting and reporting diverse characteristics of events across both IPv4 and IPv6 address space.  In our continuing work, we plan to keep a log of outage events that will be made available to the community and to refine the event reporting capabilities toward the goal of making BigBen a resource for broadly understanding Internet behavior on a daily basis.

\vspace{-0.4cm}
\section{Architecture and Implementation}\label{sec:methods}\vspace{-0.2cm}
In this section we describe the design objectives and architecture of BigBen.  We also describe the implementation that we developed to demonstrate the efficacy and utility of the system~\footnote{All software developed in this work plus a sample NTP data set and the 5 month event data set will be made public when this paper is published.}.

\begin{figure*}[htb!]
\centering
\epsfig{figure=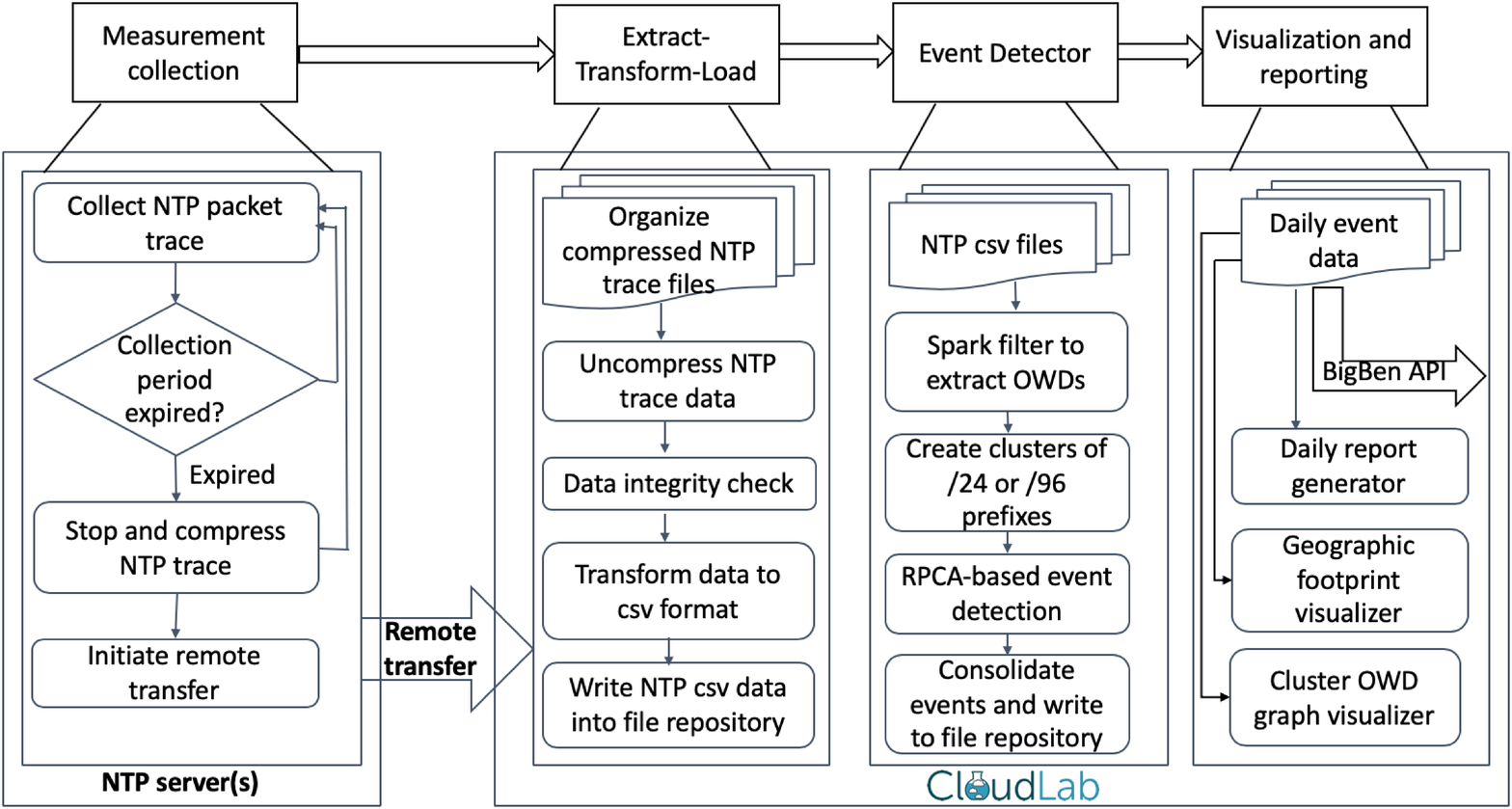,width=15cm}%
\vspace{-0.1in}
\caption{\label{fig:systemArch}{{\bf BigBen's system architecture.  The measurement component resides on each remote NTP server that contributes data, while the remaining components operate in a cloud infrastructure.}}}
\vspace{-0.2in}
\end{figure*}

\vspace{-0.3cm}
\subsection{Overview and objectives}
\label{subsection:overview}

The starting point for our work is the observation that NTP is a unique and compelling source of Internet-wide measurement data that can be readily applied to the problem of event detection~\cite{Timeweaver18,Tezzeract18}.  In particular, an Internet-wide event detection system based on NTP data offers the following advantages over active probe-based detection systems:
\begin{itemize}
\itemsep0em 
\item Easier to manage since distributed probe sources are not required,
\item Not subject to traffic blocking,
\item Doesn't introduce additional traffic into the network,
\item Fine-grained, one way delay measurements.
\end{itemize}
Another attractive aspect of NTP is that the scope and geographic diversity of clients that synchronize with any given server can be very broad~\cite{Durairajan2015HotNets}.  This offers the opportunity to detect Internet-wide events with data contributions from a relatively small number of NTP servers.  Thus, the basic model for our system is a set of distributed NTP servers that contribute data to a centralized processing system that manages the data, applies detection algorithms and generates reports on events.

The primary objective for our system is to support {\em accurate} and {\em timely} reporting of Internet events.  Accuracy is defined in the standard way in terms of low false identification rates (our system parameter configuration is conservative and aims toward erring on the side of false negatives).  Accuracy depends intrinsically on the data and algorithms applied to it.  While algorithms can be assessed using synthetic data, it is difficult to measure accuracy in practice due to the lack of outage reporting by service providers or other entities.  Partnerships with service providers are an appealing possibility.  We posit that a useful way to gain perspective on accuracy is by comparing events detected by entirely different systems.  This is exactly what we do in our evaluation of BigBen in Section~\ref{sec:results}.
Timeliness is defined from the practical perspective of being able to produce detection reports at a rate that is appropriate for a given use case such as network operations where an Internet-wide perspective would be a useful complement to internal monitoring systems.  We posit that hourly event reporting is appropriate for this use case, but for the purpose of results described in Section~\ref{sec:results}, we use daily reporting.  An important implication for timely reporting is the ability to {\em scalably and efficiently} process data from participating NTP servers.  Finally, {\em modularity} is another important objective of the design, enabling algorithms within components to be updated or replaced.  In the implementation described below, any component can be replaced and any step within a component can be reconfigured or replaced with minimal impact on other steps or components.

The design of BigBen includes four major components as shown in Figure~\ref{fig:systemArch}.  The measurement component is deployed on participating NTP servers, and the other components are deployed on a cloud-based processing infrastructure.  We describe each of the components in detail below.

\vspace{-0.3cm}
\subsection{Measurement collection component}
\label{subsection:Measurement}

\subsubsection{Process}

The objective of this component is to collect and transfer details of interactions between NTP clients and participating servers.  Standard NTP implementations do not inherently provide a logging function that collects detailed data on every connection with a client, so a customized capability is required.  The primary requirements are that the software that implements this component is easy to install and operate, that it operate securely and robustly, and that it have minimal processing/storage/network demands.  

Data collection is organized into discrete time epoch to ensure that the sizes of trace files remain modest for transfer and processing.  At the beginning of each time epoch, a program is initiated to reset trace collection to a new file, and to compress the file that was just completed. Once compressed, the trace is transferred into a cloud repository via {\tt scp} for further processing, as we discuss below.  Old traces are removed from the server to conserve disk space.


\vspace{-0.3cm}
\subsubsection{Implementation and settings}

This component is implemented in about 150 lines of Python code. 
Once initiated, this component runs constantly on the NTP server, unless it is terminated explicitly by the NTP operator or due to undesirable conditions such as network outages, server hardware failures, and so on. Each instance requires several configuration parameters, including the time period over which to create new packet trace files, and authentication credentials for data transfer using {\tt scp}.   The basis for measurements is the standard tcpdump utility, which is used to collect packet traces on server port 123 (NTP) and write the trace data to a libpcap-format file.  

We asked NTP operators who were contributing data to set the time period interval to 1 hour, to ensure modest file sizes. We note that the server contributing data for our tests with the heaviest load generates about 1.2GB compressed trace per hour on average.  We found 1 hour to be a reasonable configuration of the time period because it ensures that the remote transfer step does not overload either the network or server disk usage.

\vspace{-0.4cm}
\subsection{Cloudlab infrastructure}

The remaining data processing elements of BigBen are run on cloud-based infrastructure.  This enables high bandwidth network access, and significant processing and storage capability.  Our implementation of BigBen runs in Cloudlab~\cite{Cloudlab}. We employ a single node with the following configuration: {\em (i)} CPU: Two Intel E5-2660 v3 10-core CPUs at 2.60 GHz, and {\em (ii)} RAM: 160GB ECC memory.  Cloudlab instances typically offer two configurable 1.2 TB HDDs. Since our data storage requirement far exceeds the default available disk storage space, we employ an external 28 TB NFS disk storage. 

\vspace{-0.3cm}
\subsection{Extract-Transform-Load (ETL) component}

\subsubsection{Process}

ETL is a standard component in many commercial big data processing environments.  The basic notion is to collect and combine data from multiple sources and to place it in a target data store.  In our context, the objective of the ETL component is to receive and organize the raw NTP data from participating servers, and to prepare it for further processing.  

The steps in BigBen's ETL component include: {\em (i)} check the receipt of the compressed trace once every time epoch and to send email updates if trace files are missing;  {\em (ii)} organize the compressed trace files by month for each NTP server; {\em (iii)} extract relevant NTP information from the compressed libpcap traces; {\em (iv)} assure data integrity by removing empty or corrupted records; and {\em (v)} transform the extracted NTP information from the files to a CSV format for further processing.  This last step enables us to obtain one-way delay (OWD) information from the client to the NTP server and vice-versa (from server to client). 

This component makes our data processing pipeline more robust by identifying and reporting interruptions in the receipt of the compressed traces to the NTP server operators.  In practice this feature proved very useful in interactions with the operators and re-initiation the measurement collection component (see Section~\ref{subsection:Measurement}) on multiple occasions. For example, during the month of February 2019, we did not receive data for the majority of days from NTP servers C1 to C4, E1, and E2 (described in Section~\ref{section:Datasets}) due to the fact that the NTP operator had to seek additional permission to run BigBen over the long term.  Once the required permission was obtained, the data collection process was resumed.

\vspace{-0.3cm}
\subsubsection{Implementation and settings}

The high-level steps of this component are implemented in about 180 lines of Bash code. This script is scheduled to run as a cron job once during every time epoch (1 hour - coordinated with measurement collection). This component takes less than a minute for all 16 servers (see Section~\ref{section:Datasets}). The transformation step which converts tcpdump data to CSV format is implemented in about 700 lines of C code. For every packet in the tcpdump data, CSV schema includes the following fields: {\em (1)} packet number, {\em (2)}, source IP address, {\em (3)} destination IP address, {\em (4)} latency (NTP origin timestamp - NTP receive timestamp), {\em (5)} NTP polling exponent, {\em (6)} packet timestamp, {\em (7)} root delay, {\em (8)} round-trip time, and {\em (9)} NTP reference IP address. 

\vspace{-0.3cm}
\subsection{Event identification component}
\label{subsection:Detector}

The objective of this component is to generate a set of {\em events} that constitute a {\em significant} deviation from normal behavior for a given network prefix.  This definition aligns directly with a standard definition of anomaly detection.  In our context of NTP-based monitoring, it relates to the idea of identifying a significant change in OWD between one or more clients in a given network prefix and a participating NTP server.

The high level steps involved in the event identification component are as follows.  First, we process the CSV data to identify OWDs from clients that are in {\em tight synchronization} with the NTP server.  Next, we cluster clients into IPv4 /24 or IPv6 /96 prefixes (selected as a convention for detection and reporting).  We then employ an event detector based on Robust Principle Components Analysis (RPCA) to identify delay spikes experienced by clients within a cluster.  We apply a classifier based on Z-scores to label events from high to low confidence.  Finally, we consolidate events to provide a single, global view of events. We describe the details of these steps below.

\vspace{-0.4cm}
\subsubsection{Extracting OWDs}

A filtering step is required to extract OWDs from trace data.  In particular, we want to use OWDs for event detection from clients that are in tight synchronization ({\em i.e.,} the NTP exchange has been observed to be stable over a period of time) with their server since these delays are more likely to have smaller error bounds~\cite{Timeweaver18}.

The key consideration for this step is the large volume of data processing required ({\em e.g.}, about 72 GB for one particular server) for 24 hours.  While the measurement and the ETL components execute hourly, the event identification component is run on a daily basis.  The reason we chose to run this component every 24h is to ensure that there are ample data points for the latency filtering algorithm described below.  However, this processing window is configurable and we have the capability to execute this component for a shorter processing time window. 

\textit{Implementation and settings}. We implemented OWD filtering in Apache Spark~\cite{spark} to parallelize CSV trace processing from a single NTP server.  The filter is written in pyspark.  It reads the CSV data as Resilient Distributed Dataset (RDD) chunks in a stand-alone mode, single node Spark cluster\footnote{This is not an inherent limitation, but simply reflects our CloudLab setup.}. We process only the client-to-server packets since they provide OWD measurement values in both directions---latency computation (field 4 of CSV data) provides server to client (s2c) OWD, and root delay (field 7 of CSV data) provides client to server (c2s) OWD.  We then process the RDD to aggregate per-client c2s and s2c OWDs. The aggregation step is non-deterministic in pyspark, that is, the order of aggregated OWDs need not correspond to the temporal ordering of the packets. So, after the aggregation step, we use the packet timestamp to sort and re-establish the temporal packet ordering for each client's aggregated data. 

Next, we identify any trend in changes to the NTP polling value for each client and classify the clients into constant, increasing, decreasing, and variable polling types. For each classification type, we implement the specific filtering algorithm described in~\cite{Timeweaver18} to identify tightly synchronized (TS) clients. These OWD values associated with these clients are then written into RDD-specific chunk files. The pyspark processing for the server with heaviest data load takes about 50 minutes.  Next, we cluster clients into /24 and /96 prefixes for IPv4 and IPv6 respectively. The pyspark implementation and the clustering implementation is executed by a wrapper Bash script (about 290 lines of code), which executes filtering and clustering for each NTP server sequentially. The execution time for processing 24-hour data (about 100 GB) for all 16 servers (see Section~\ref{section:Datasets}) is about 1 hour and 15 minutes.  For future work, we intend to parallelize filtering and clustering across all NTP server inputs.

\vspace{-0.4cm}
\subsubsection{RPCA event detection}
\label{subsubsection:RPCA}

We use an RPCA-based method to identify events in the form of OWD spikes experienced by a cluster of clients. RPCA was shown to be an effective technique for processing NTP data to detect events in~\cite{Tezzeract18}.   The modular design of BigBen enables other detectors to be substituted or used in parallel.  Our implementation of the RPCA algorithm is shown in Algorithm~\ref{alg:rpcaEventDetector}. For each cluster, we generate a $t \times n$ matrix similar to~\cite{Tezzeract18} (steps 1 to 21), where $t$ is the time bin used to group every client in a row and $n$ is the number of clients in a prefix cluster.  Instead of using pcaNA estimation for the NA values in the matrix, we replace the NA values for each client with the minimum latency for the client. To ensure that this doesn't hinder spike detection, we compared the RPCA scores for a cluster using synthetic events with both pcaNA estimation and minimum latency estimation, and found the RPCA results to be similar in terms of the detected outliers. We employ the Robust Covariance Estimator (RCE)~\cite{croux2000principal} for the clusters, by default (steps 22 to 32).  We found that nearly 50\% of the clusters have a covariance matrix with non-negative definite, hence RCE fails on such clusters. For the clusters on which RCE fails, we employ the Elliptical PCA (EPCA)~\cite{locantore1999robust} method to detect outliers. Again, using a cluster with known events, we compared RCE and EPCA results to confirm that RPCA scores are similar for both of these methods. 

\textit{Implementation and settings}.   Our RPCA-based detector is implemented in about 1000 lines of Python code.  The implementation is parallelized for a single NTP server's filtered OWD data. We employ Pebble~\cite{pebble} for multiprocess generation of the $t \times n$ clusters, invocation of RPCA (we leverage PcaNA R package~\cite{pcana}), and generation of events by aggregating contiguous timestamp bins with RPCA outliers---false flag values (steps 33 and 34). If an event is detected in a single timestamp bin we classify it as {\em single spike}.  If an event is detected in multiple timestamp bins, it is classified as an {\em event}.


\vspace{-0.4cm}
\subsubsection{Z-score event classification}
\label{subsection:classification}

To aid in assuring internal consistency and to provide a measure of confidence in detected events and single spikes we employ a Z-score computation.  For every $t \times n$ cluster, we generate Z-scores on OWDs to identify outliers defined as 2$\sigma$ deviation from $\mu$.  We then  classify events as follows (steps 35 and 36 in Algorithm~\ref{alg:rpcaEventDetector}): {\em (1)} class A: 75\% to 100\% correlation between RPCA outliers and Z-score outliers (these are the highest confidence events), {\em (2)},  class B: 50\% to 75\% correlation, {\em (3)} class C: 25\% to 50\% correlation, and {\em (4)} class D: 0\% to 25\% correlation. The implementation for the classification is part of the event detector code. Due to very minimal correlation with Z-score outliers, we discard all the events classified into class D, and further process only events from Class A, B, and C for our system evaluation (see Section~\ref{sec:results}).

The event identification component generates two primary output files. The first is the event details output file with entries consisting of prefix cluster, number of clients, event start time, event end time, and event class details for every event. The second is the matrix details output file consisting of all the $t \times n$ matrices for the prefix clusters, which forms the input to BigBen's visualization component (see Section~\ref{subsubsection:clusterViz}).

\vspace{-0.4cm}
\subsubsection{Event consolidation}
\label{subsect:consolidation}

The event consolidation step post-processes the events detected with the individual 16 NTP servers trace datasets, and consolidates them to provide a single global view of events.   The consolidation process begins by taking the events detected using one server as the \textit{base set}. We then employ the following steps in order to merge the remaining server events: {\em (i)} obtain the list of matching /24 or /96 prefix clusters and the corresponding events {\em (ii)} employ Python's intervaltree package~\cite{intervaltree} to determine direct temporal overlap between previously observed events on the same prefix, and  {\em (iii)} merge the events which have overlap by considering the earliest event start time, latest event end time, and best possible class between class A, class B, and class C (see Section~\ref{subsection:classification}).  While there are other ways to perform event consolidation, we adopt a simple approach and note that the modular design of our system enables us to replace this step with other methods, if needed.  At the conclusion of this step the raw /24 and /96 event data and the consolidated raw event data are output to files.

\vspace{-0.3cm}
\subsection{Visualization and reporting component}
\label{subsection:visualization}

A significant amount of raw event data can be generated by BigBen.  This calls for tools for post processing that can be used for organizing, analyzing and understanding Internet-wide events on a daily or hourly basis.  This includes reporting of simple statistics and providing visualizations that enable insights on aspects of events including network aggregates, geographic location, etc.  

\vspace{-0.3cm}
\subsubsection{Daily report generator}
The daily report generator provides simple statistics on events detected over the previous 24 hours.  This includes information about BigBen itself including the total amount of data contributed by participating servers, total size of the daily CSV files, total number of clients, etc.  Daily reports also include the total number of /24 and /96 prefixes in which events were detected, the top 10 largest events in terms of larger network aggregates, the top 10 longest duration events and other basic information.  The daily reports are useful for monitoring BigBen itself and for developing a general understanding of network event behaviors.

\subsubsection{Geographic footprint visualizer}
\label{subsubsection:bigfootViz}
The geographic footprint visualizer component enables visualization of the geolocation of IPv4 clusters (network prefixes) for which events are detected. Given a network prefix, this component geolocates each IP address using using MaxMind's IP Geolocation service~\cite{maxmind}. Then, it plots the geolocated points on a map and creates a convex hull of the geolocated points. A similar approach is reported in~\cite{syamkumar16}.  For /24 prefixes, typically this component generates a visualization of a circle around the single geographic location to which the IP addresses are geolocated. The backend processing for this component is implemented in about 500 lines of Python code and the frontend visualizer is built using ESRI ArcGIS~\cite{arcgis}.

\vspace{-0.3cm}
\subsubsection{Cluster OWD graph visualizer}
\label{subsubsection:clusterViz}
The cluster OWD graph visualizer component produces one-way delay timeseries graphs for an IPv4 or IPv6 cluster with detected events. This component is implemented in about 150 lines of Python code. To handle the varying number of clients per prefix, this component uses seaborn~\cite{seaborn} to generate a lineplot for each client and automatically assign color mapping to the clients using the hue parameter. Before plotting the timeseries, we replace the NA values for each client with the minimum latency for the client, similar to the RPCA processing (see Section~\ref{subsubsection:RPCA}).

\begin{algorithm}[htb!]
\SetKwInOut{Input}{input}
\newcommand{\cmtsty}[1]{\texttt{\small #1}}\SetCommentSty{cmtsty}
\Input{OWDs from TS clients}
\Input{$prefixClusters$}
\ForEach{prefix $P$ in prefixClusters}{
	$leastPollValues$ = [];\\
	\ForEach {client $C$ in $P$}{
		$pollValues$ = getPollFromPacket($C$);\\
		$leastPollValue$ = min($pollValues$);\\
		$leastPollValues$.append($leastPollValue$);
	}
	$t$ = median($leastPollValues$);\\
	$n$ = len($P$);\\
	$sEpoch$, $eEpoch$ = getEpochs();\\
	$timeBins$ = generateTimeBins($sEpoch$, $eEpoch$, $t$);\\
	$t*n_{matrix}$ = generateEmptyMatrix($t$, $n$);\\
	\ForEach {client $C$, $C_{index}$ in $P$}{
		$measurements$ = getMeasurements($C$);\\
		\ForEach {$m$ in $measurements$}{
			$owd$, $epochTS$ = getDetails($m$);\\
			$t_{index}$ = findTimeBin($epochTS$, $timeBins$);\\
			$pOwd$ = getValue($t*n_{matrix}$, $t_{index}$, $C_{index}$);\\
			\If{$owd$ > $pOwd$}{
				\tcp{retain maximum OWD per bin}
				setValue($t*n_{matrix}$, $t_{index}$, $C_{index}$, $owd$);
			}
		}
	}
	$t'*n_{matrix}$ = removeAllNARows($t*n_{matrix}$);\\
	$t'*n_{matrix}$ = setMinOWDValues($t'*n_{matrix}$);\\
	$eigenValues$ = generateEigenValues($t'*n_{matrix}$);\\
	$sum$ = sum($eigenValues$);\\
	$variancePercentages$ = [];\\
	\ForEach {$e$ in $eigenValues$}{
		$percentage$ = ($e$ / $sum$) * 100;\\
		$variancePercentages$.append($percentage$);
	}
	$top_k$ = 0;\\
	\ForEach {$p$ in $variancePercentages $}{
		\If{$p$ > 5}{
			$top_k$++;
		}
	}
	$results$ = pcaNA($t'*n_{matrix}$, $scale$ = True, $k$ = $top_k$);\\
	$flags$ = getScoreFlags($results$);\\
	\tcp{False score indicates anomaly}
	\tcp{Consecutive occurrence of False flag constitutes one event}
	$ePrefix$, $eStart$, $eEnd$ = getEvents($flags$, $timeBins$);\\
	$zScoreOutliers$ = computeZScoreOutliers($t'*n_{matrix}$);\\
	$eventClass$ = classifyEvents($flags$, $zScoreOutliers$);
}
\caption{BigBen EventDetector}
\label{alg:rpcaEventDetector}
\end{algorithm}

\vspace{-0.3cm}
\section{NTP Datasets}\label{sec:data}\vspace{-0.3cm}
\label{section:Datasets}

In this section, we provide an overview of the NTP trace datasets collected from 16 NTP servers distributed across 7 geographic sites. The geographic sites include: {\em (i)} Chicago, IL (C), {\em (ii)} Edison, NJ (E), {\em (iii)} Jackson, WI (J), {\em (iv)} Philadelphia, PA (P), {\em (v)} Salt Lake City, UT (S), {\em (iv)} Urbana-Champaign, IL (U), and {\em (vii)} Madison, WI (M).

Table~\ref{tab:statsBasic} summarizes the basic statistics from each NTP server for the months of January to May 2019. A total of 15.5TB raw NTP trace dataset was collected during this time period.   As shown in the table, the amount of data collected from the servers varies significantly, with server S1 generating 3.2GB per hour on an average, which is much larger when compared to the datasets from the rest of the NTP servers. The big data processing requirement (100GB per day on an average) was a key consideration in BigBen's system design and implementation.  Furthermore, we note that the majority of the NTP servers support both IPv4 and IPv6 clients; both protocols are also supported in BigBen's event detector component (see Section~\ref{subsection:Detector}). To the best of our knowledge, our event detection system is the first to provide complete support for detecting events in the IPv6 address space.
  
 Data provided during the study was fairly consistent.  However, there were days in which we did not receive data from individual providers.   Reasons for gaps in the dataset include: downtime of NTP server, outage of Cloudlab network, and NFS network downtime.  Overall, this represented a small fraction of the total period of collection.  

The coverage and accuracy of detected events relates directly to servers providing data to BigBen.  As will be shown in Section~\ref{sec:results}, our data set includes clients from only a subset of the v4 address space.  However, we argue that the volume and coverage is sufficient for demonstrating the capabilities BigBen and to show how NTP-based monitoring provides a useful, complementary perspective on events.  We plan  to expand the number of NTP servers monitored by BigBen in future work.



\begin{table}[t!]
\centering
  {\caption{\label{tab:statsBasic}{{\bf Summary of NTP traces collected by BigBen during the period of January 2019 to May 2019.}}}}
  \vspace{0.1in}
\begin{small}
\begin{tabular}{ |p{2cm}|c|c|r|r| }
\hline
Server	&	Server	& IP	&	\multicolumn{1}{c|}{Raw data}	&	\multicolumn{1}{c|}{csv data} 	\\
Location	&	ID	&  Version	&	\multicolumn{1}{c|}{size}	&	\multicolumn{1}{c|}{size} 	\\ \hline \hline
\multirow{4}{*}{\centering Chicago, IL} & C1	& v4/v6	
		&		13.3G	&	12.38G		\\ \cline{2-5}
& C2	& v4/v6 &		13.61G	&	12.38G		\\ \cline{2-5}
& C3	& v4/v6 &		16.4G	&	15.17G		\\ \cline{2-5}
& C4	& v4/v6 &		11.76G	&	10.83G		\\ \hline
\multirow{2}{*}{\centering Edison, NJ} & E1	& v4/v6	
		&		11.14G	&	10.52G		\\ \cline{2-5}
& E2	& v4/v6 &		11.45G	&	10.52G		\\ \hline
\multirow{3}{*}{\centering Jackson, WI} & J1	& v4	
		&		13.26G	&	10.76G		\\ \cline{2-5}
& J2	& v4 &		65.09G	&	65.09G		\\ \cline{2-5}
& J3	& v4	&		11.09G		&	10.74G		\\ \hline
{\centering Philadelphia, PA} & P1	& v4/v6
		&		6.84G	&	6.18G		\\ \hline
{\centering Salt Lake City, UT} & S1	& v4/v6
		&		10.76T	&	10.04T		\\ \hline
\multirow{3}{*}{\parbox{2cm}{Urbana-Champaign, IL}} & U1	& v4/v6	
		&		282.3G	&	192.48G		\\ \cline{2-5}
& U2 & v4/v6 &		306.12G	&	208.57G		\\ \cline{2-5}
& U3 & v4/v6 &		189.86G	&	128.72G		\\ \hline
\multirow{2}{*}{\centering Madison, WI} & M1	& v4	
		&		3.73T	&	3.73T		\\ \cline{2-5}
& M2	 & v4 &		100.94G	&	104.55G		\\ \hline
\end{tabular}
\vspace{-0.2in}
\end{small}
\end{table}

\vspace{-0.3cm}
\section{System Evaluation}\label{sec:results}In this section, we demonstrate BigBen's ability to detect events. We begin by reporting the basic characteristics of the events detected in our NTP data set.  Then, we compare the events detected by our methodology with the events detected through active probing from ISI's Trinocular project~\cite{Heidemann08a_200802}. Finally, we report on the perspective that BigBen provides on events that were reported by third parties, highlighting the detailed results from one specific event. 

\vspace{-0.3cm}
\subsection{Basic event characteristics}

\begin{figure*}[htb!]
\centering
\vspace{+0.02in}
\epsfig{figure=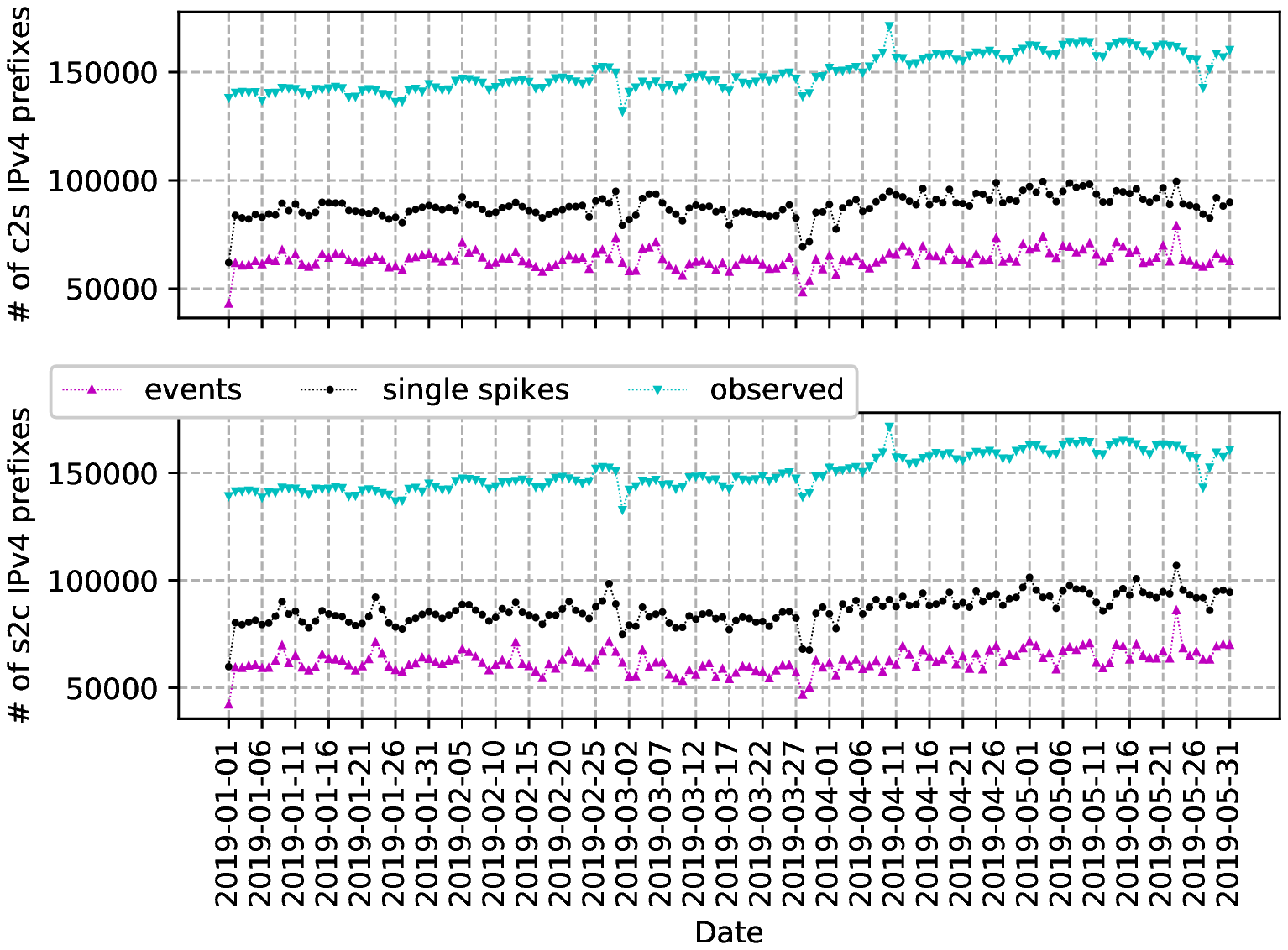,width=8.8cm}
\epsfig{figure=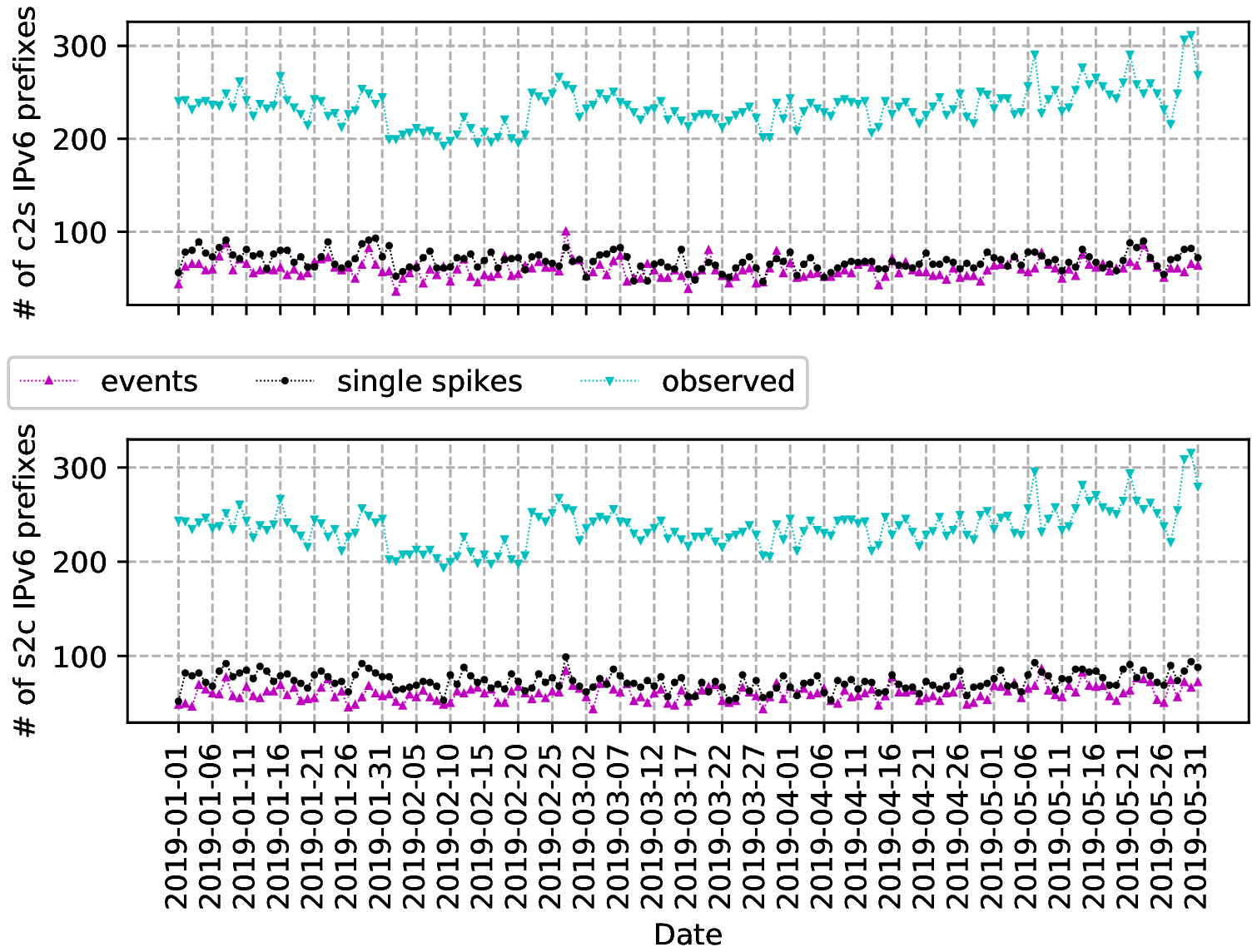,width=8.8cm}
\vspace{-0.2in}
\caption{\label{fig:prefixCount}{{\bf Number of IPv4 /24 (left) and IPv6 /96 (right) prefixes observed and number of events detected within those prefixes over the 5 months period of study.}}}
\vspace{-0.2in}
\end{figure*}

Figure~\ref{fig:prefixCount}-(left) shows the total number of IPv4 prefix clusters observed in BigBen dataset, and total number of IPv4 prefixes in which events were detected from January 2019 to May 2019. It also shows the number of IPv4 prefixes on which single spikes were detected. The scope of the covered IPv4 prefixes ranges from 2 clients to 244 clients. Figure~\ref{fig:prefixCount}-(right) shows similar results for IPv6 prefixes, whose scope of coverage ranges from 2 clients to 84 clients. These figures show that the number of prefixes on which BigBen detects events is quite consistent on a daily basis and has a slightly increasing trend over the period of study.  The basic characteristics of observed prefixes observed and events detected is relatively equivalent from both the c2s and s2c perspective. 

The duration of the detected c2s IPv4 events ranges from 32 seconds to 27 hours, with an average event duration of 2.7 hours. For IPv4 s2c events, the duration range remains the same, with an average event duration of 2.6 hours. For IPv6 c2s events, event durations range from 64 seconds to 24 hours, with an average event duration of 3.5 hours. For IPv6 s2c events, the average event duration is 1.7 hours. IPv4 and IPv6 single spikes (both s2c and c2s) have a very short lived durations, ranging from 1 second to 22 minutes (due to clients with high NTP polling values). 

\begin{figure*}[htb!]
\centering
\vspace{+0.02in}
\epsfig{figure=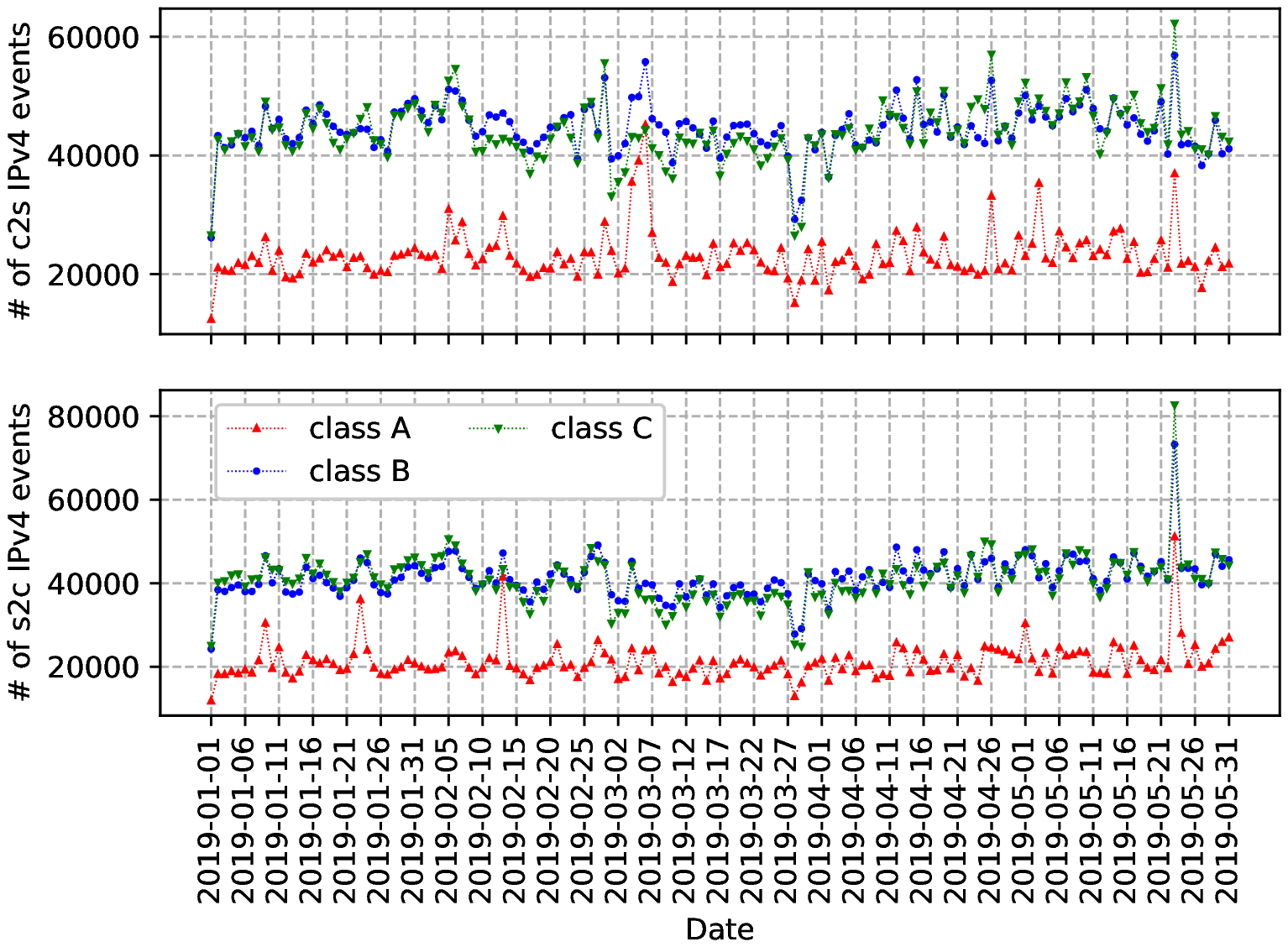,width=8.8cm}
\epsfig{figure=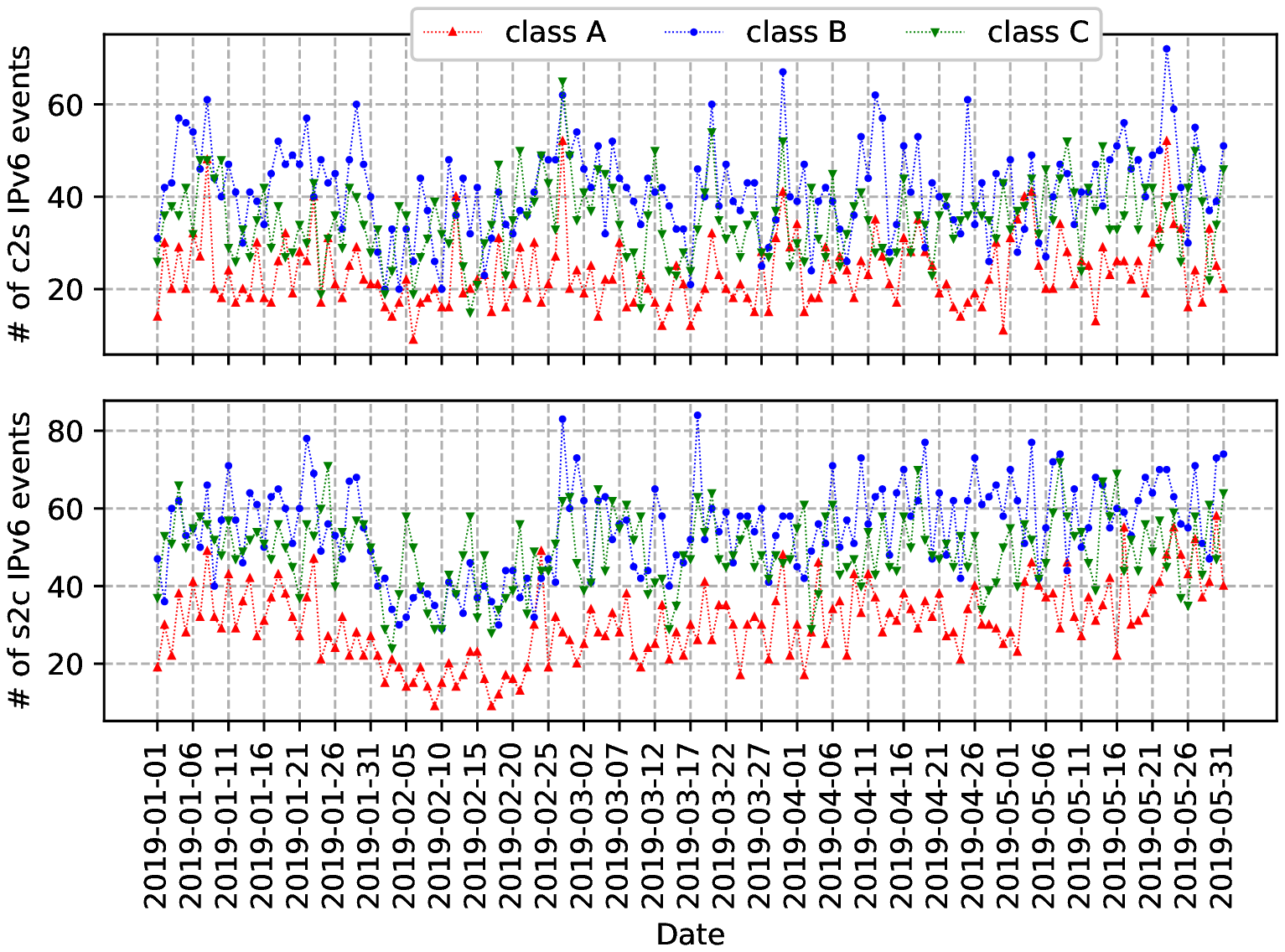,width=8.8cm}
\vspace{-0.2in}
\caption{\label{fig:eventsCount}{{\bf Number of events detected for IPv4 /24 (left) and IPv6 /96 (right) prefixes broken down by event classes as described in Section~\ref{subsection:classification}.}}}
\vspace{-0.1in}
\end{figure*}

Figure~\ref{fig:eventsCount} shows details of the timeseries of IPv4 events (left) and IPv6 events detected by BigBen in our data set. The figure shows some consistency in terms of number of events identified by BigBen on a daily basis.  We posit that many of these reflect standard internet dynamics such as route changes, congestion or address block reconfiguration.  We also observe that there are fewer high-confidence (class A) than lower-confidence events (classes B and C) over time.  Interestingly, the spikes observed in the timeseries ({\em e.g.}, May 23) do not correspond to reported events from third parties; we are investigating these events in our ongoing work.  When considering similar timeseries of single spike events (not shown due to space constraints), we (a) observe a similar pattern of fewer high-confidence events (class A) than lower-confidence events, as well as (b) a very high count of single spikes (as expected) due to the permissive nature of spike identification.


\vspace{-0.3cm}
\subsection{Event consolidation and aggregation across prefixes}
\vspace{-0.1cm}

The consolidation process is applied individually to events and single spikes ({\em c.f.}~\ref{subsubsection:RPCA}). On a daily average, we are able to consolidate 103 IPv4 events, 9 IPv6 events, 82 IPv4 single spikes, 5 IPv6 single spikes detected on c2s OWDs, and 114 IPv4 events, 2 IPv6 events, 102 IPv4 single spikes, 5 IPv6 single spikes detected on s2c OWDs. We expected the count of events observed across servers to be low, as we initially consider clusters of small-sized prefixes (/24 or /96), showing that additional NTP servers will further diversified and enhance coverage. 

\begin{figure*}[htb!]
\centering
\vspace{+0.02in}
\epsfig{figure=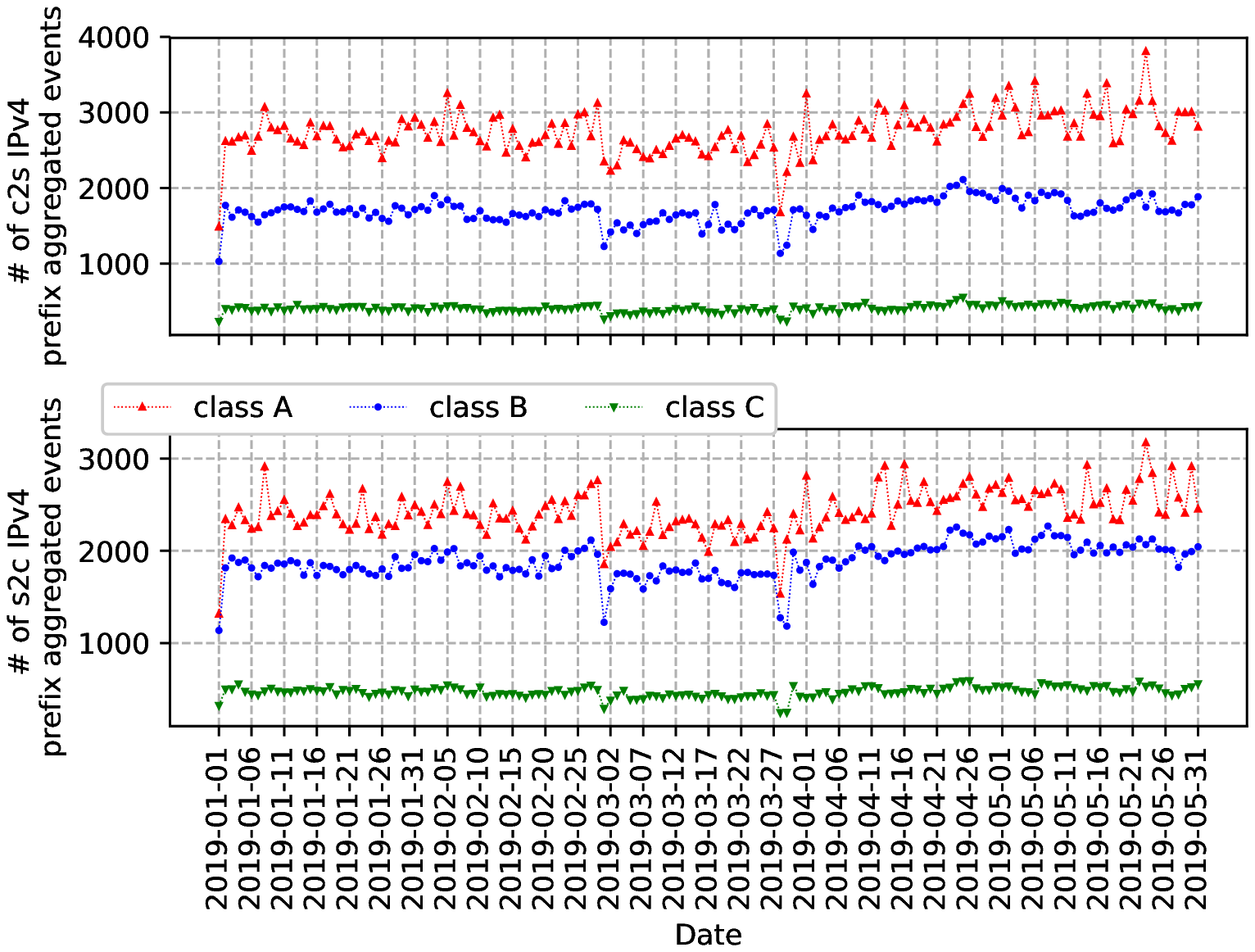,width=8.5cm}
\epsfig{figure=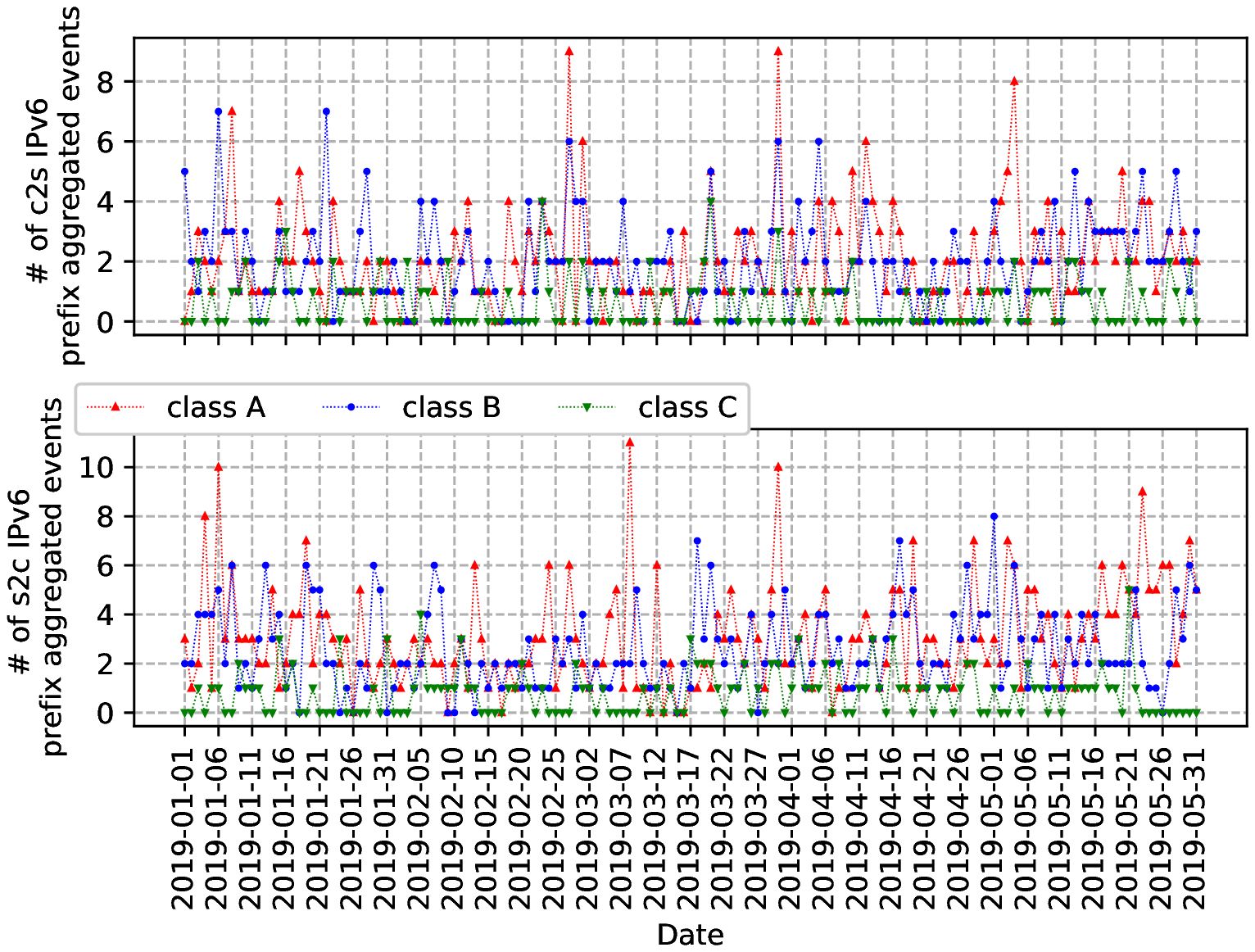,width=8.5cm}
\vspace{-0.2in}
\caption{\label{fig:caidaCount}{{\bf Number of IPv4 (left) and IPv6 (right) events detected by announced network prefix based on prefix lists provided by CAIDA and Team Cymru during the period of study.}}}
\vspace{-0.2in}
\end{figure*}

The event aggregation post-processes the consolidated events (see Section~\ref{subsect:consolidation}) detected on /24 or /96 prefixes into larger network aggregates ({\em i.e.,} shorter prefixes). To perform this aggregation, we consider prefix lists from CAIDA~\cite{pfx2as} and Team Cymru~\cite{cymru} and look up each /24 or /96 prefix to find the corresponding larger IP prefix aggregate, typically included in BGP announcements. For every higher level IP prefix where we observe at least 2 constituent /24 or /96 prefixes, we aggregate the events using the same steps as the consolidation algorithm (see Section~\ref{subsect:consolidation}). For this aggregation process, we only consider the BigBen events and omit BigBen single spikes. Note also that the confidence level assigned to an event is the \textit{highest} observed across the events that get aggregated together.  Figure~\ref{fig:caidaCount} shows the count of aggregated IPv4 events (left) and IPv6 events (right). As shown in the figure, the aggregation process is more significant for IPv4 prefixes---aggregation impacts nearly 5000 events on a per-day basis for both c2s and s2c events. This aggregation process ensures that we don't repetitively count the events belonging to the same higher level IP prefix, while generating a ranking of top 5 Autonomous Systems (AS) on which we observe events. 

Table~\ref{tab:rankingASv4} shows the ranking for top 5 IPv4 ASes in terms of identified events, including AS number, name, count of BigBen events observed throughout the 5 month dataset. Interestingly, we observe a large number of events for Amazon, Inc. We show that the same Amazon AS suffers collateral damage during a outage event reported by a third party (see Section~\ref{subsect:groundTruthEvents}). BigBen also identifies a sizable number of events in tier-1 service providers ({\em e.g.,} Verizon, CenturyLink). Table~\ref{tab:rankingASv6} shows a similar ranking for top 5 IPv6 ASes.

\vspace{-0.3cm}
\begin{table}[htb!]
\centering
  {\caption{\label{tab:rankingASv4}{{\bf Top 5 ASes based on the number of IPv4 events identified from Jan-May 2019.}}}}
\begin{small}
\begin{tabular}{ |c|l|l|l|l }
\hline
Rank	&	ASN	&	ISP	&	\# of events		\\	\hline \hline
1    &   16509 & Amazon.com &  565,523     \\
2    &   22773 & Cox Communications & 536,379       \\
3    &   20115 & Charter Communications & 527,132       \\
4    &   701 & Verizon Business  &    342,186     \\
5    &   209 & CenturyLink    &    258,535  \\
    &    & Communications   &   \\ \hline
\end{tabular}
\end{small}
\end{table}

\begin{table}[htb!]
\centering
  {\caption{\label{tab:rankingASv6}{{\bf Top 5 ASes ranked based on the number of IPv6 events identified from Jan-May 2019.}}}}
\begin{small}
\begin{tabular}{ |c|l|l|l|l }
\hline
Rank	&	ASN	&	ISP	&	\# of events	\\	\hline \hline
1    &   32748 & Steadfast &  8,631     \\
2    &   14061 & DigitalOcean & 2,871      \\
3    &   26347 & New Dream Network & 1,792       \\
4    &   7922 & Comcast Cable  &    1,435     \\
    &   & Communications  &     \\
5    &   54825 & Packet Host   &    1,390   \\ \hline
\end{tabular}
\vspace{-0.15in}
\end{small}
\end{table}

\vspace{-0.1in}
\subsection{Comparison with ISI/Trinocular}

Next, we compare the IPv4 events detected by BigBen with the events detected by ISI's Trinocular project~\cite{Heidemann08a_200802}.  The ISI project regularly probes addresses on all reachable IPv4 /24's (minus those that have requested no probing). For this comparison, we consider the ISI dataset from January to March 2019. During this period, the ISI dataset represented a total of 4,226,299 IPv4 prefixes.  BigBen's dataset over the same period included clients from 488,269 IPv4 prefixes.  Interestingly, 76,545 IPv4 prefixes are observed in our data that are {\em not} observed by ISI.  This highlights one of the complementary aspects of our NTP-based approach. During the comparison period, ISI reports an average of 121,695 IPv4 events on a per-day basis, while BigBen reports an average of 99,878 c2s events and 109,761 s2c events on a per-day basis.  

\begin{figure*}[htb!]
\centering
\vspace{+0.02in}
\epsfig{figure=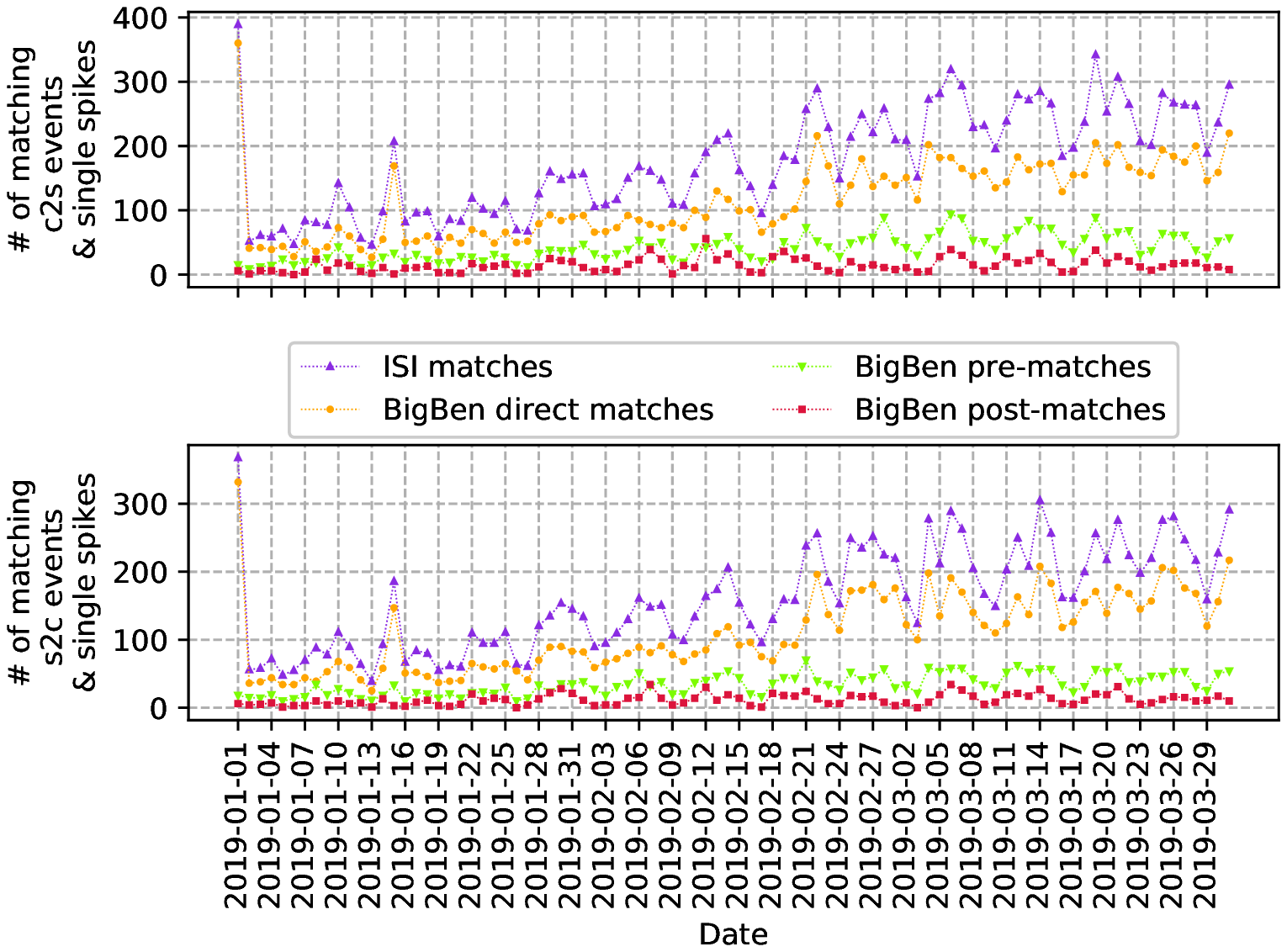,width=8.8cm}
\epsfig{figure=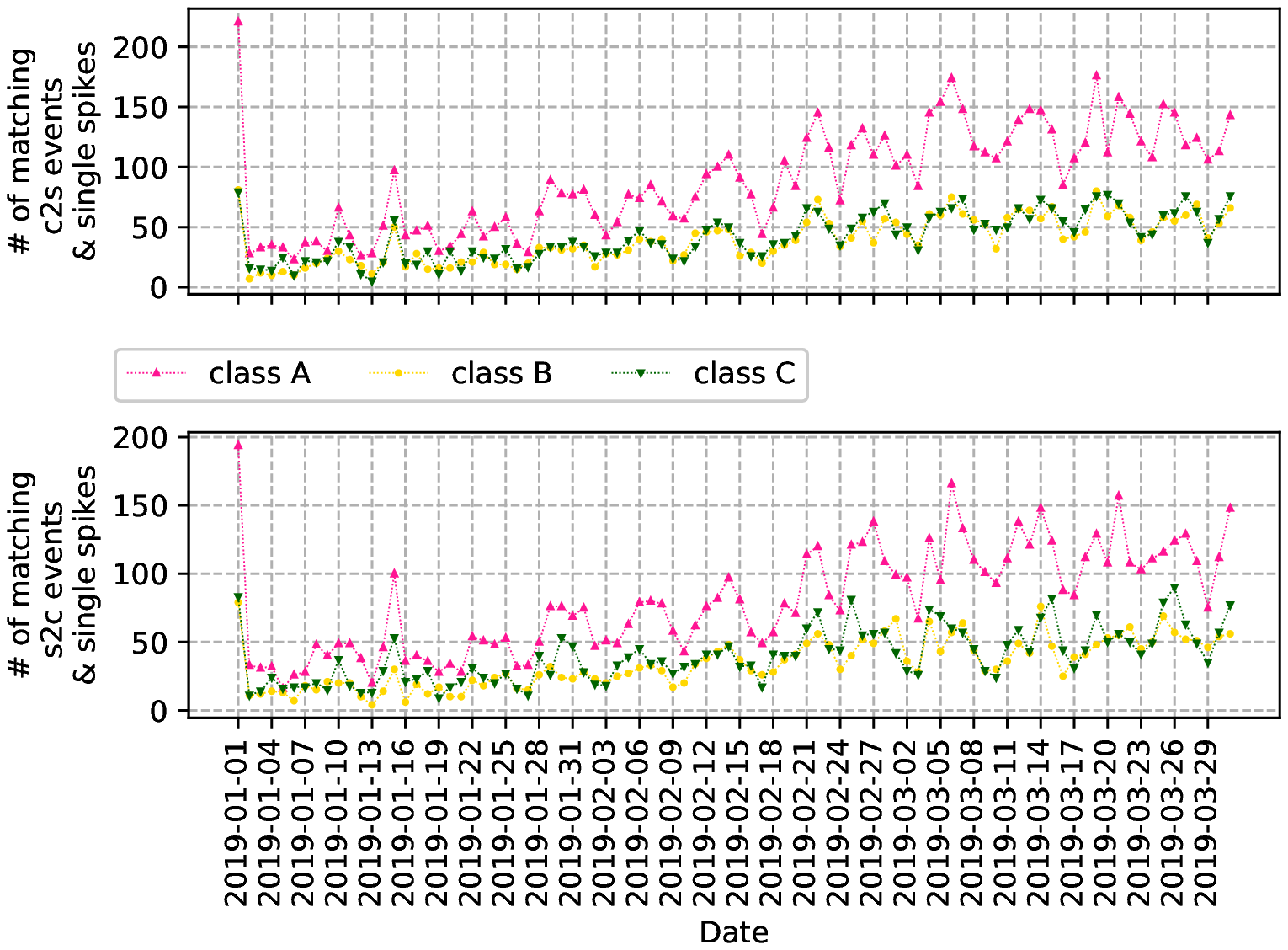,width=8.8cm}
\vspace{-0.2in}
\caption{\label{fig:isiMatches}{{\bf Number of ISI/Trinocular event matches (left) and corresponding classes for matching BigBen events (right).}}}
\vspace{-0.2in}
\end{figure*}

\vspace{-0.3cm}
\subsubsection{Event matching}

We match events detected by BigBen with ISI-reported events based on /24 prefixes using a similar approach to our event consolidation (see Section~\ref{subsect:consolidation}). Apart from looking for a direct temporal overlap between BigBen events and ISI events we also consider events that match within a time window, which we set to be 1 hour. We include both events and single spikes for this comparison. Figure~\ref{fig:isiMatches}-(left) shows the number of BigBen events and single spikes that have direct matches, pre-matches, and post-matches with ISI events on a per day basis. It also shows the corresponding number of matching ISI events. We attribute the relatively low event match counts to the fact that we observe a similarly low match in /24 prefixes with events between the two datasets---only 40,097 matches (for the entire 3 month dataset). Figure~\ref{fig:isiMatches}-(right) shows the classes to which the matching BigBen events and single spikes belong to. As expected, we see higher number of matching events fall under class A (BigBen's highest confidence level class). 

\vspace{-0.4cm}
\subsection{Events reported by third parties}
\label{subsect:groundTruthEvents}

We compare BigBen detected events to a set of events reported by third party sources including CAIDA, the Outages mailing list, and ThousandEyes, Inc.  We consider these third-party reported events a form of ground truth and useful for demonstrating the utility of NTP-based event detection.  It is worth mentioning that when we generated a list of the largest events (terms of total number of /24 prefixes affected) detected by BigBen in  over the period of study, none overlapped with the third-party reported events.  We are conducting forensic investigation of these and other events in on-going work.

Table~\ref{tab:knownEvents} shows a summary of reported events found during our data collection period. For every event, we report the event start and end time observed by the third party and the corresponding start and end times observed by BigBen. We provide a list of Autonomous Systems (AS) impacted during the event and the count of /24 prefixes on which BigBen observes the event.  We also list the geographical region of impact of the event, obtained by visualizing the event using our Geographic footprint visualizer (see~\ref{subsubsection:bigfootViz}). These results highlight BigBen's capabilities to detect events across geographical regions throughout the world. We show visualizations from one of the third party reported events below and omit the rest due space limitations.  

\begin{table*}[htb!]
\centering
  {\caption{\label{tab:knownEvents}{{\bf Comparison of BigBen detected events with reported outage events.}}}}
\begin{small}
\begin{tabular}{|l|l|l|l|l|l|l|l|}
\hline
\begin{tabular}[c]{@{}l@{}}External\\ event\\ start time\end{tabular} & \begin{tabular}[c]{@{}l@{}}External\\ event\\ end time\end{tabular} & \begin{tabular}[c]{@{}l@{}}BigBen\\ events\\ start time\end{tabular} & \begin{tabular}[c]{@{}l@{}}BigBen\\ events\\ end time\end{tabular} & \begin{tabular}[c]{@{}l@{}}BigBen - \\ AS(es)\\ affected\end{tabular} & \begin{tabular}[c]{@{}l@{}}BigBen - \\ \# of /24 \\ prefixes \\ affected\end{tabular} & \begin{tabular}[c]{@{}l@{}}BigBen - \\ countries\\ affected\end{tabular} & External event source \\ \hline \hline
Jan 20, 	&	Jan 20,	&	Jan 20, 	&	Jan 20, 	&	AS18809,		&	87		&	Panama,			&	CAIDA~\cite{jan20CAIDA} \\ 
4:28 PM	&	7:08 PM	&	3:30 PM	&	8:04 PM	&	AS20473		&			&	USA				&	\\ \hline
Jan 25, 	&	Jan 25, 	&	Jan 25,  	&	Jan 25,  	&	AS7018		&	437		&	USA				&	Outages mailing list ~\cite{jan25OutageMailingList}, \\
4:28 PM 	&	6:48 PM	&	3:28 PM	&	7:47 PM	&				&			&					&	Verizon network status ~\cite{jan25Verizon}  \\ \hline
Jan 29,  	& 	Jan 29,  	& 	Jan 29,  	& 	Jan 29, 	& 	AS18809,   	& 	184 		& 	Panama,  			& 	CAIDA ~\cite{jan29CAIDA} \\
2:47 PM	&	9:40 PM	&	1:48 PM	&	10:37 PM	&	AS22205,  	&			&	USA				&	\\
		&			&			&			&	AS43350, 		&			&					&	\\
		&			&			&			&	AS20473		&			&					&	\\ \hline
Feb 21,  	& 	Feb 21, 	& 	Feb 21,	&	Feb 21,	&	AS327712 	& 	1		& 	Algeria			&	CAIDA ~\cite{feb21CAIDA}, \\ 
7:49 PM	&	11:09 PM	&	8:06 PM 	&	8:07 PM 	&				&			&					&	NetBlocks ~\cite{feb21NetBlocks} \\ \hline
Feb 26,   	& 	Feb 27,  	& 	Feb 26,   	&	Feb 27,  	& 	AS6400, 		&	15		& 	Dominican		& 	CAIDA ~\cite{feb26CAIDA} \\
2:00 PM	&	3:15 AM	&	1:49 PM	&	2:36 AM	&	AS51964		&			&	Republic, 			&	\\ 
		&			&			&			&				&			&	Germany			&	\\ \hline
Mar 7,  	& 	Mar 13,  	& 	Mar 08,	& 	Mar 13, 	& 	AS8048, 		&  	5		&	Venezuela, 		&	CAIDA ~\cite{mar7CAIDA}, \\
12:50 PM	&	8:30 PM	&	10:29 AM	&	6:41 PM	&	AS21826,  	&			&	Germany,			&	Miami Herald ~\cite{mar7MiamiHerald}, \\
		&			&			&			&	AS51964,		&			&	USA				&	Havana Times ~\cite{mar7HavanaTimes} \\ 
		&			&			&			&	AS11878,		&			&					&	\\ \hline
Apr 28,	& 	Apr 29, 	& 	Apr 28,  	& 	Apr 28,  	& 	AS36907 		& 	1		& 	Angola			& 	CAIDA ~\cite{apr28CAIDA} \\
10:00 AM &	6:00 AM	&	11:14 AM	&	11:20 AM	&				&			&					&	\\ \hline
April 29,  	& 	May 1,  	& 	Apr 29,  	& 	May 1,  	& 	AS3491 		& 	4		&	UK, USA			& 	CAIDA ~\cite{apr29CAIDA} \\
12:30 PM	&	4:00 AM	&	3:29 PM	&	1:43 AM	&				&			&	Greece			&	\\ \hline
May 13,  	& 	May 14,  	& 	May 13,  	&	May 14, 	& 	AS16591,  	& 	7176 	&	USA, 			& 	CAIDA ~\cite{CAIDAChineseEvent}, \\
7:30 PM	&	12:30 AM	&	6:32 PM	&	1:26 AM 	&	AS56042		&			&	China			&	ThousandEyes ~\cite{thousandEyesChineseEvent}, \\
		&			&			&		 	&	AS16509		&			&					&	Internet Disruption Report ~\cite{IDRChineseEvent} \\ 
		&			&			&			&	AS38895		&			&					&	\\ \hline
May 13,  	& 	May 14,  	& 	May 13, 	& 	May 14,  	& 	AS63008,   	& 	1174 	&	USA				& 	CAIDA ~\cite{may13CAIDA} \\
11:30 PM	&	9:00 AM	&	10:33 PM	&	9:54 AM	&	AS209		&			&					&	\\ \hline
May 18, 	& 	May 18,  	& 	May 18,  	& 	May 18,  	& 	AS237,  		& 	15		& 	USA				& 	CAIDA ~\cite{may18CAIDA} \\
12:30 AM 	&	11:43 PM	&	12:00 AM	&	11:27 PM	&	AS7377, 		&			&					&	\\
		&			&			&			&	AS3651		&			&					&	\\ \hline
May 19,  	& 	May 19,  	& 	May 19,  	& 	May 19,  	& 	AS63008, 		& 	4 		& 	USA				& 	CAIDA ~\cite{may19CAIDA} \\
8:30 AM	&	9:00 PM	&	10:12 AM	&	8:36 PM	&	AS11878		&			&					&	\\ \hline
\end{tabular}
\vspace{-0.2in}
\end{small}
\end{table*}

\vspace{-0.3cm}
\subsubsection{May 13, 2019 China Telecom Event}
\begin{figure}[htb!]
 \centering
 \epsfig{figure=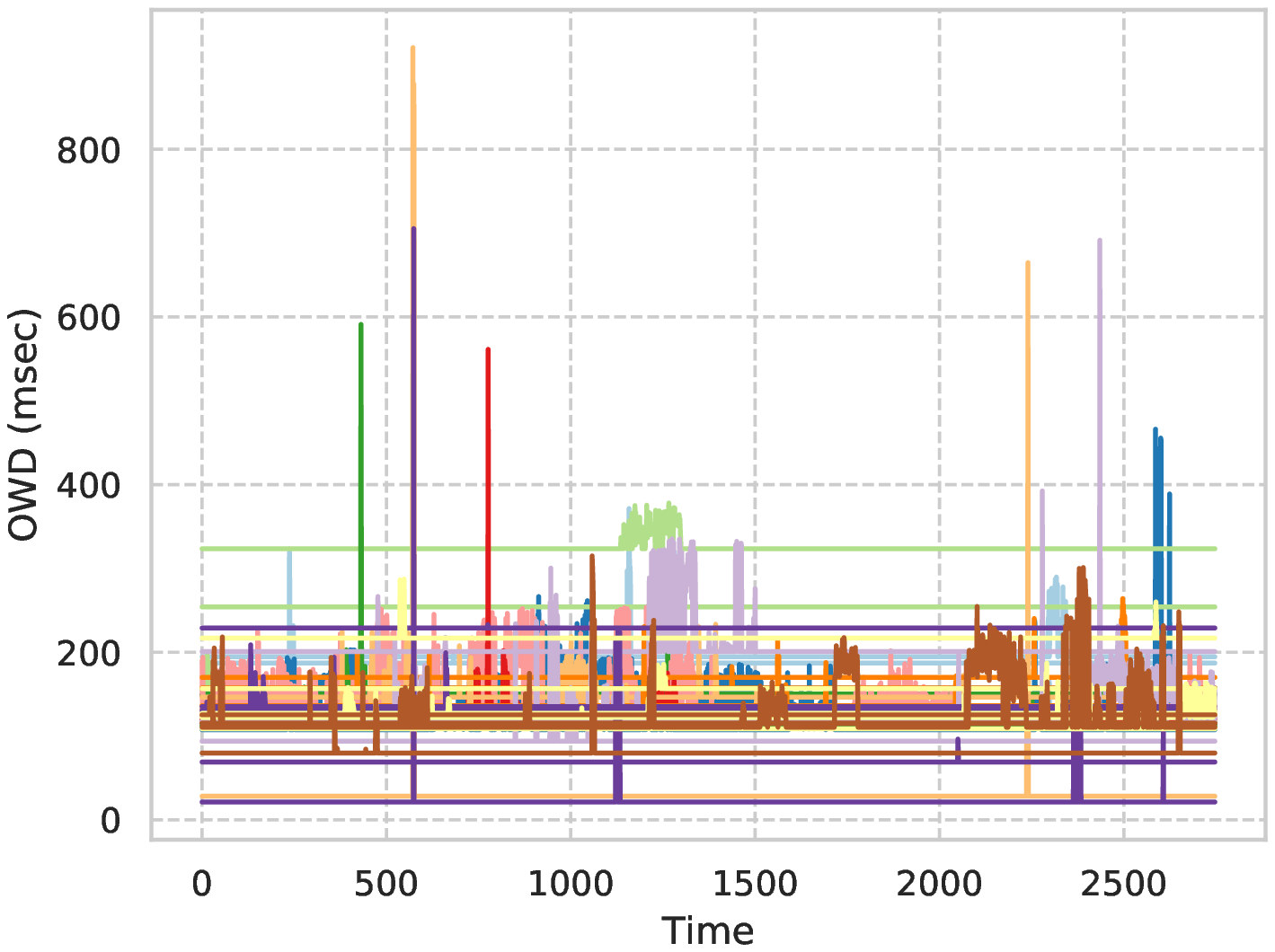,width=8.8cm}
 \vspace{-0.2in}
 \caption{\label{fig:chinaEventOWD}{{\bf May 13, 2019 China Telecom outage OWD graph for clusters 117.136.4.0/24.}}}
 \vspace{-0.1in}
 \end{figure}

\begin{figure}[htb!]
 \centering
 \epsfig{figure=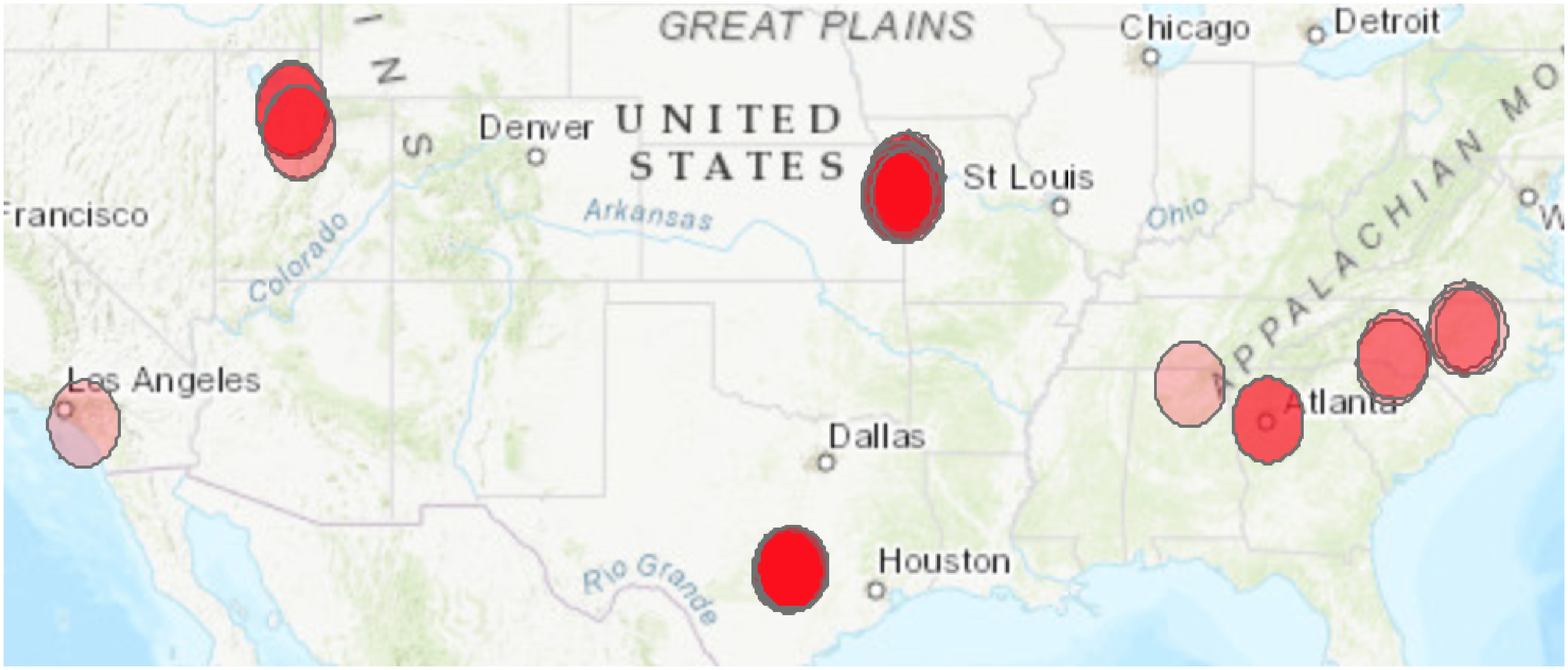,width=8.8cm}
 \vspace{-0.2in}
 \caption{\label{fig:chinaEventBigfoot}{{\bf May 13, 2019 China Telecom outage geographic footprints of impacted prefixes in USA.}}}
 \vspace{-0.2in}
 \end{figure}

ThousandEyes reported~\cite{thousandEyesChineseEvent} that on May 13, 2019, China Telecom suffered a significant network outage lasting about 5 hours. BigBen's reporting showed significant OWD elevations on multiple prefixes during this time. Figure~\ref{fig:chinaEventOWD} shows the BigBen's Cluster OWD graph visualization (c2s OWDs) for prefix cluster 117.136.4.0/24 affecting 59 clients. This prefix is allocated to AS56042 (China Mobile Communications Corp.). In the figure, we observe delay spikes to nearly 900 milliseconds, indicating a 4x increase in OWD.  These observations are consistent with those reported by ThousandEyes. In BigBen's geographical impact plot, we observe effects across various parts of the United States (see Figure~\ref{fig:chinaEventBigfoot}), as well as in China (plot not shown due to space considerations).  While ThousandEyes reported that the US West Coast was affected, BigBen's report shows that the impact was more widespread across the US. Consistent with ThousandEyes reporting, BigBen's report also observes Amazon AS16509 and AS38895 suffering collateral damage due to this outage event.


\vspace{-0.4cm}
\section{Related Work}\label{sec:relwork}\vspace{-0.3cm}
There have been many prior studies of detecting, locating, and diagnosing network outages.  Similar to our work, a number of these previous studies have relied on streams of \textit{passively collected data} to infer failures and outages.  For example, some works have used feeds of routing updates from BGP or from other routing protocols, {\em e.g.}, ~\cite{labovitz1998internet,labovitz1999experimental, markopoulou2008characterization, pucha2007understanding}.  Others have analyzed router BGP configurations to detect or better understand the causes of certain types of failures~\cite{mahajan2002understanding,feamster2005detecting}.  Router logs (syslogs) have also been a key data source for analyzing the nature of network anomalies and faults~\cite{lakhina2004diagnosing,qiu2010happened}.  The potential synergy of combining routing updates with syslog data has been examined by Turner {\em et al.}~\cite{turner2013comparison}, and Roughan {\em et al.} used the same data sources to detect forwarding anomalies which may be indicative of misconfigurations or impending failures~\cite{roughan2004combining}.  Additional work in the area of network fault detection has evaluated communications among network operators to analyze network failures~\cite{banerjee2015internet}.  Turner {\em et al.} used syslog data, email announcements, and router configurations in an effort to understand causes behind network failures~\cite{turner2010california}.  Network service providers are in a privileged position to collect a great deal of passive data for understanding the nature and causes of network faults; Markopolou {\em et al.} reported on findings from the Sprint IP backbone~\cite{markopoulou2008characterization}.  Other studies over the years have used passively collected data from network honeypots to to identify both malicious and benign Internet events~\cite{pang04,benson2013gaining}.  Studies have also used ever-present network protocol traffic~\cite{zhu2012latlong} to identify network prefixes where faults may be taking place through detection of changes in delays or the absence of ordinarily-present traffic.  

Our work is most similar to prior studies on NTP traffic and the opportunity to utilize this traffic to detect network events~\cite{Durairajan2015HotNets,Timeweaver18,Tezzeract18}.  Indeed, we build directly on that body of work.  However, our contributions are in the design and implementation of a scalable NTP data processing system that can be used on a daily basis for internet-wide event monitoring and easily modified to utilize a variety of detection methods.

Another category of studies related to ours are those have focused on network-wide monitoring and event detection within large data center infrastructures.  Data center monitoring challenges are similar to those that we faced in development of BigBen including managing vary large data sets and identifying events accurately and in a timely fashion.  Systems for data center monitoring and event detection are often based on passive measurements collected on end hosts or switches.  For example, Moshref {\em et al.} describe Trumpet, which is a system designed to provide high precision data center-wide monitoring on end hosts~\cite{Moshref16}.  Similarly, techniques for {\em software defined measurement} on programmable switches have been proposed in~\cite{Jose11,Moshref14,Yu13}.  The primary difference between these studies and ours is our focus on {\em Internet-wide} monitoring, which includes the challenge of gathering data from diverse networks. 

Many studies, in contrast to those mentioned so far, have used \textit{active measurement} to identify and localize network outages and impairments.  RFC 2678 identifies metrics for network \textit{connectivity} and \textit{reachability}, which relate directly to the notion of network faults and disconnection~\cite{rfc2678}.  Many prior works have used the basic notion of RFC 2678---that a packet sent to a network address receives a response within some finite amount of time---as the basis for identifying network faults.  For example, the Trinocular project uses periodic pings at a low rate to the full IPv4 address space, and detect disruptions and outages through a probabilistic framework~\cite{quan2013trinocular,quan2013towards,quan2012detecting}.  As noted above, we use some of the data from this project as a point of comparison.  Along the same lines, Padmanabhan {\em et al.} evaluated the response times to pings across the IPv4 address space and found that a small fraction of responses (5\%) take at least 5 seconds to arrive~\cite{padmanabhan2015timeouts}.  This result has significant implications for the design of probe-based systems for detecting outages.  The work by Schulman {\em et al.} used weather reports as a trigger to send targeted probes for detecting weather-related network outages or failures~\cite{schulman2011pingin}.  Probe-based techniques have also been widely used to detect routing loops and other anomalies, path failures, and elevated end-to-end delay or abrupt changes in delay~\cite{paxson1997end,paxsonthesis,roughan2004combining,wang2006measurement,pucha2007understanding}.  Moreover, \textit{tomographic} methods and similar types of algorithms have been created to both detect and localize faults to particular links or portions of a path~\cite{duffield2006network,dhamdhere2007netdiagnoser,huang2008practical,bejerano2006robust,barford2009network,kompella2007detection}.  Finally, similar to our work, there are a few systems that are designed to identify network failures and other events on an ongoing basis.  For example, the Planetseer system was a Planetlab-based system designed to detect wide-area network failures~\cite{zhang2004planetseer}.  Similarly, the iPlane system also utilized the geographic distribution of Planetlab hosts for ongoing detection of network reachability problems~\cite{madhyastha2006iplane}, and the Hubble system used both periodic active measurements and passive analysis of BGP updates to trigger on-demand traceroutes for verifying and localizing network outages~\cite{Katz08}.  We believe that active probe-based methods and systems for Internet event identification will continue to provide useful insights, and that BigBen provides a complementary perspective that will enhance and enrich future analyses of events.

\vspace{-0.4cm}
\section{Conclusions}\label{sec:conclusions}\vspace{-0.3cm}
Fiber cuts, malicious attacks, configuration changes, censorship and persistent congestion are examples of the {\em events} that routinely degrade or disrupt service in the Internet.  Identifying and characterizing the details of such events can lead to new methods, configurations, protocols and systems that improve Internet service.  This paper describes {\em BigBen}, a cloud-based network telemetry processing system designed to support accurate and timely detection of events throughout the Internet.   
BigBen's design includes {\em (i)} distributed collection of passive measurements of NTP traffic, {\em (ii)} an Extract Transform Load component that organizes data into a normalized format, {\em (iii)} an event identification component, and {\em (iv)} a visualization and reporting component.  We develop a cloud-based implementation of BigBen that it is able to process tarabytes of data on a daily basis.  We demonstrate this implementation on a 15.5TB corpus of NTP data collected over a period of five months.  The results show a wide range of events characterized by their scope and duration. We compare the events detected by BigBen vs. events detected by a large active probe-based detection system.  We find only modest overlap and show how BigBen provides details on events that are not available from active measurements. We also show how BigBen identifies outage events that have been more broadly reported by third parties.  Next steps for BigBen include expanding partnerships with NTP servers to broaden reach and depth of event identification, expanding visualizations and reporting, and to develop methods for forensic analysis and classification of events. 
\bibliographystyle{plain}
\bibliography{paper}

\begin{thebibliography}{10}

\bibitem{spark}
{Apache Spark.}
\newblock \url{https://spark.apache.org/}, 2019.

\bibitem{apr28CAIDA}
{CAIDA April 28, 2019 to April 29, 2019 outage.}
\newblock
  \url{https://ioda.caida.org/ioda/dashboard\#view=inspect\&entity=country/BJ\&lastView=overview\&from=1556385008\&until=1556557928},
  2019.

\bibitem{apr29CAIDA}
{CAIDA April 29, 2019 to May 1, 2019 outage.}
\newblock
  \url{https://ioda.caida.org/ioda/dashboard\#from=1556518575\&until=1556864175\&lastView=overview\&view=inspect\&entity=country/NE},
  2019.

\bibitem{feb21CAIDA}
{CAIDA February 21, 2019 outage.}
\newblock
  \url{https://ioda.caida.org/ioda/dashboard\#view=inspect\&entity=country/DZ\&lastView=overview\&from=1550736000\&until=1550822400},
  2019.

\bibitem{feb26CAIDA}
{CAIDA February 26, 2019 outage.}
\newblock
  \url{https://ioda.caida.org/ioda/dashboard\#view=inspect\&entity=country/DO\&lastView=overview\&from=1551168000\&until=1551254400},
  2019.

\bibitem{jan29CAIDA}
{CAIDA Jan 29, 2019 outage.}
\newblock
  \url{https://ioda.caida.org/ioda/dashboard\#view=inspect\&entity=country/PA\&lastView=overview\&from=1548748800\&until=1548835200},
  2019.

\bibitem{jan20CAIDA}
{CAIDA January 20, 2019 outage.}
\newblock
  \url{https://ioda.caida.org/ioda/dashboard\#from=1547964000\&until=1548050400\&view=inspect\&entity=country/PA\&lastView=overview},
  2019.

\bibitem{mar7CAIDA}
{CAIDA March 7, 2019 to March 13, 2019 outage.}
\newblock
  \url{https://ioda.caida.org/ioda/dashboard\#view=inspect\&entity=country/VE\&lastView=overview\&from=1551946200\&until=1552528800},
  2019.

\bibitem{CAIDAChineseEvent}
{CAIDA May 13, 2019 outage.}
\newblock
  \url{https://ioda.caida.org/ioda/dashboard\#view=inspect\&entity=country/CN\&lastView=overview\&from=1557724024\&until=1557810424},
  2019.

\bibitem{may13CAIDA}
{CAIDA May 13, 2019 to May 14, 2019 outage.}
\newblock
  \url{https://ioda.caida.org/ioda/dashboard\#from=1557642244\&until=1558074244\&lastView=overview\&view=inspect\&entity=country/NE},
  2019.

\bibitem{may18CAIDA}
{CAIDA May 18, 2019 outage.}
\newblock
  \url{https://ioda.caida.org/ioda/dashboard\#from=1558134164\&until=1558223024},
  2019.

\bibitem{may19CAIDA}
{CAIDA May 19, 2019 outage.}
\newblock
  \url{https://ioda.caida.org/ioda/dashboard\#from=1558163974\&until=1558423174\&lastView=overview\&view=inspect\&entity=country/CM},
  2019.

\bibitem{pfx2as}
{CAIDA Routeviews Prefix to AS mappings Dataset (pfx2as) for IPv4 and IPv6.}
\newblock \url{https://www.caida.org/data/routing/routeviews-prefix2as.xml},
  2019.

\bibitem{Cloudlab}
{Cloudlab platform.}
\newblock \url{https://cloudlab.us/}, 2019.

\bibitem{arcgis}
{ESRI ArcGIS.}
\newblock \url{https://www.esri.com/en-us/arcgis/about-arcgis/overview}, 2019.

\bibitem{mar7HavanaTimes}
{Havana Times - Internet Outage for 96\% of Venezuela with Prolonged
  Blackouts.}
\newblock
  \url{https://havanatimes.org/uncategorized/internet-outage-for-96-of-venezuela-with-prolonged-blackouts/},
  2019.

\bibitem{thousandEyesChineseEvent}
{Internet Outage Reveals Reach of China's Connectivity.}
\newblock
  \url{https://blog.thousandeyes.com/internet-outage-reveals-reach-of-chinas-connectivity/},
  2019.

\bibitem{maxmind}
{MaxMind IP Geolocation Service.}
\newblock \url{https://www.maxmind.com/}, 2019.

\bibitem{mar7MiamiHerald}
{Miami Herald - Venezuela\'s power outage threatens information blackout as
  internet collapses.}
\newblock
  \url{https://www.miamiherald.com/news/nation-world/world/americas/venezuela/article227296244.html},
  2019.

\bibitem{IDRChineseEvent}
{Multi-Hour Disruption at China Telecom.}
\newblock
  \url{https://internetdisruption.report/2019/05/16/multihour-disruption-at-china-telecom/},
  2019.

\bibitem{feb21NetBlocks}
{NetBlocks - Multiple targeted internet disruptions in Algeria amid
  mass-demonstrations.}
\newblock
  \url{https://netblocks.org/reports/algeria-internet-disruptions-amid-mass-demonstrations-WJBZjMB6},
  2019.

\bibitem{jan25OutageMailingList}
{[Outages mailing list]. AT\&T/SBC Global issues}.
\newblock
  \url{https://puck.nether.net/pipermail/outages/2019-January/012036.html},
  2019.

\bibitem{intervaltree}
{Python package: A mutable, self-balancing interval tree for Python 2 and 3.}
\newblock \url{https://pypi.org/project/intervaltree/}, 2019.

\bibitem{pebble}
{Python package: Pebble - Threading and multiprocessing eye-candy.}
\newblock \url{https://pypi.org/project/Pebble/}, 2019.

\bibitem{pcana}
{R package: Classical Or Robust Principal Components For Incomplete Data.}
\newblock
  \url{https://www.rdocumentation.org/packages/rrcovNA/versions/0.4-8/topics/PcaNA},
  2019.

\bibitem{seaborn}
{seaborn: statistical data visualization.}
\newblock \url{https://seaborn.pydata.org/}, 2019.

\bibitem{cymru}
{Team CYMRU's IP to ASN mapping.}
\newblock \url{http://www.team-cymru.org/IP-ASN-mapping.html}, 2019.

\bibitem{jan25Verizon}
{Verizon - Health check: Network status}.
\newblock
  \url{https://status.verizondigitalmedia.com/pages/incident/5736344c90417cda1a000f3f/5c4b4c4023395904bd439c78},
  2019.

\bibitem{banerjee2015internet}
Ritwik Banerjee, Abbas Razaghpanah, Luis Chiang, Akassh Mishra, Vyas Sekar,
  Yejin Choi, and Phillipa Gill.
\newblock {Internet outages, the eyewitness accounts: Analysis of the outages
  mailing list}.
\newblock In {\em PAM}, 2015.

\bibitem{barford2009network}
Paul Barford, Nick Duffield, Amos Ron, and Joel Sommers.
\newblock Network performance anomaly detection and localization.
\newblock In {\em IEEE INFOCOM}, 2009.

\bibitem{bejerano2006robust}
Yigal Bejerano and Rajeev Rastogi.
\newblock {Robust monitoring of link delays and faults in IP networks}.
\newblock {\em IEEE/ACM Transactions On Networking}, 14(5):1092--1103, 2006.

\bibitem{benson2013gaining}
Karyn Benson, Alberto Dainotti, Kimberly~C Claffy, and Emile Aben.
\newblock {Gaining insight into AS-level outages through analysis of Internet
  background radiation}.
\newblock In {\em IEEE INFOCOM workshops}, 2013.

\bibitem{croux2000principal}
{Croux, Christophe and Haesbroeck, Gentiane}.
\newblock {Principal Component Analysis based on Robust Estimators of the
  Covariance or Correlation matrix: Influence Functions and Efficiencies}.
\newblock {\em Biometrika}, 2000.

\bibitem{dhamdhere2007netdiagnoser}
Amogh Dhamdhere, Renata Teixeira, Constantine Dovrolis, and Christophe Diot.
\newblock {Netdiagnoser: Troubleshooting network unreachabilities using
  end-to-end probes and routing data}.
\newblock In {\em ACM CoNEXT conference}, 2007.

\bibitem{duffield2006network}
Nick Duffield.
\newblock Network tomography of binary network performance characteristics.
\newblock {\em IEEE Transactions on Information Theory}, 52(12):5373--5388,
  2006.

\bibitem{Timeweaver18}
R.~Durairajan, S.~Mani, P.~Barford, R.~Nowak, and J.~Sommers.
\newblock {TimeWeaver: Opportunistic One Way Delay Measurement via NTP}.
\newblock In {\em {Proceedings of ITC30 - Teletraffic in a Smart World}},
  September 2018.

\bibitem{feamster2005detecting}
Nick Feamster and Hari Balakrishnan.
\newblock {Detecting BGP configuration faults with static analysis}.
\newblock In {\em Proceedings of NSDI}, pages 43--56, 2005.

\bibitem{Sonata18}
A.~Gupta, R.~Harrison, M.~Canini, N.~Feamster, J.~Rexford, and W.~Willinger.
\newblock {Sonata: query-driven streaming network telemetry}.
\newblock In {\em {Proceedings of the AMC SIGCOMM Conference}}, August 2018.

\bibitem{Heidemann08a_200802}
John Heidemann, Yuri Pradkin, Ramesh Govindan, Christos Papadopoulos, Genevieve
  Bartlett, and Joseph Bannister.
\newblock {Census and Survey of the Visible Internet (extended)}.
\newblock {\em ISI-TR-2008-649}, 2008.

\bibitem{huang2008practical}
Yiyi Huang, Nick Feamster, and Renata Teixeira.
\newblock Practical issues with using network tomography for fault diagnosis.
\newblock {\em ACM SIGCOMM Computer Communication Review}, 38(5):53--58, 2008.

\bibitem{Jose11}
L.~Jose, M.~Yu, and J.~Rexord.
\newblock {Online Measurement of Large Traffic Aggregates on Commodity
  Switches}.
\newblock In {\em {Proceedings of Hot-ICE}}, March 2011.

\bibitem{Katz08}
Ethan Katz-Bassett, Harsha~V Madhyastha, John~P John, Arvind Krishnamurthy,
  David Wetherall, and Thomas~E Anderson.
\newblock {Studying Black Holes in the Internet with Hubble}.
\newblock In {\em NSDI}, 2008.

\bibitem{kompella2007detection}
Ramana~Rao Kompella, Jennifer Yates, Albert Greenberg, and Alex~C Snoeren.
\newblock Detection and localization of network black holes.
\newblock In {\em IEEE INFOCOM 2007}, pages 2180--2188, 2007.

\bibitem{labovitz1999experimental}
Craig Labovitz, Abha Ahuja, and Farnam Jahanian.
\newblock {Experimental study of Internet stability and backbone failures}.
\newblock In {\em International Symposium on Fault-Tolerant Computing}, 1999.

\bibitem{labovitz1998internet}
Craig Labovitz, G~Robert Malan, and Farnam Jahanian.
\newblock Internet routing instability.
\newblock {\em IEEE/ACM Transactions on Networking}, 1998.

\bibitem{lakhina2004diagnosing}
Anukool Lakhina, Mark Crovella, and Christophe Diot.
\newblock {Diagnosing Network-wide Traffic Anomalies}.
\newblock In {\em ACM SIGCOMM}, 2004.

\bibitem{locantore1999robust}
N~Locantore, JS~Marron, DG~Simpson, N~Tripoli, JT~Zhang, KL~Cohen, Graciela
  Boente, Ricardo Fraiman, Babette Brumback, Christophe Croux, et~al.
\newblock {Robust Principal Component Analysis for Functional Data}.
\newblock {\em Test}, 1999.

\bibitem{madhyastha2006iplane}
Harsha~V Madhyastha, Tomas Isdal, Michael Piatek, Colin Dixon, Thomas Anderson,
  Arvind Krishnamurthy, and Arun Venkataramani.
\newblock {iPlane: An information plane for distributed services}.
\newblock In {\em Proceedings of OSDI}, pages 367--380, 2006.

\bibitem{mahajan2002understanding}
Ratul Mahajan, David Wetherall, and Tom Anderson.
\newblock {Understanding BGP misconfiguration}.
\newblock In {\em ACM SIGCOMM Computer Communication Review}, volume~32, pages
  3--16, 2002.

\bibitem{rfc2678}
J.~Mahdavi and V.~Paxson.
\newblock {RFC 2678: IPPM Metrics for Measuring Connectivity}.
\newblock \url{https://tools.ietf.org/html/rfc2678}, September 1999.

\bibitem{markopoulou2008characterization}
Athina Markopoulou, Gianluca Iannaccone, Supratik Bhattacharyya, Chen-Nee
  Chuah, Yashar Ganjali, and Christophe Diot.
\newblock Characterization of failures in an operational ip backbone network.
\newblock {\em IEEE/ACM Transactions on Networking}, 2008.

\bibitem{Moshref14}
M.~Moshref, M.~Yu, R.~Govindan, and A.~Vahdat.
\newblock {DREAM: Dynamic Resource Allocation for Software-defined
  Measurement}.
\newblock In {\em {Proceedings of the AMC SIGCOMM Conference}}, August 2016.

\bibitem{Moshref16}
M.~Moshref, M.~Yu, R.~Govindan, and A.~Vahdat.
\newblock {Trumpet: Timely and Precise Triggers in Data Centers}.
\newblock In {\em {Proceedings of the AMC SIGCOMM Conference}}, August 2016.

\bibitem{padmanabhan2015timeouts}
Ramakrishna Padmanabhan, Patrick Owen, Aaron Schulman, and Neil Spring.
\newblock Timeouts: Beware surprisingly high delay.
\newblock In {\em ACM IMC}, 2015.

\bibitem{pang04}
R.~Pang, V.~Yegneswaran, P.~Barford, V.~Paxson, and L.~Peterson.
\newblock {Characteristics of Internet Background Radiation}.
\newblock In {\em {Proceedings of ACM Internet Measurement Conference}},
  October 2004.

\bibitem{paxsonthesis}
V.~Paxson.
\newblock {\em {Measurements and Analysis of End-to-end Internet Dynamics}}.
\newblock PhD thesis, University of California, Berkeley, 1997.

\bibitem{paxson1997end}
Vern Paxson.
\newblock {End-to-end routing behavior in the Internet}.
\newblock {\em IEEE/ACM transactions on Networking}, 1997.

\bibitem{pucha2007understanding}
Himabindu Pucha, Ying Zhang, Z~Morley Mao, and Y~Charlie Hu.
\newblock Understanding network delay changes caused by routing events.
\newblock In {\em ACM SIGMETRICS Performance Evaluation Review}. ACM, 2007.

\bibitem{qiu2010happened}
Tongqing Qiu, Zihui Ge, Dan Pei, Jia Wang, and Jun Xu.
\newblock {What happened in my network: mining network events from router
  syslogs}.
\newblock In {\em Proceedings of ACM Internet Measurement Conference}, pages
  472--484, 2010.

\bibitem{quan2012detecting}
Lin Quan, John Heidemann, and Yuri Pradkin.
\newblock {Detecting Internet outages with precise active probing (extended)}.
\newblock {\em USC/Information Sciences Institute, Tech. Rep}, 2012.

\bibitem{quan2013towards}
Lin Quan, John Heidemann, and Yuri Pradkin.
\newblock {Towards active measurements of edge network outages}.
\newblock In {\em International Conference on Passive and Active Network
  Measurement}, pages 276--279, 2013.

\bibitem{quan2013trinocular}
Lin Quan, John Heidemann, and Yuri Pradkin.
\newblock {Trinocular: Understanding Internet reliability through adaptive
  probing}.
\newblock In {\em ACM SIGCOMM Computer Communication Review}, 2013.

\bibitem{Durairajan2015HotNets}
{R. Durairajan and S. Mani and J. Sommers and P. Barford}.
\newblock {Time's Forgotten: Using NTP to Understand Internet Latency}.
\newblock In {\em ACM HotNets}, 2015.

\bibitem{roughan2004combining}
Matthew Roughan, Tim Griffin, Morley Mao, Albert Greenberg, and Brian Freeman.
\newblock {Combining routing and traffic data for detection of IP forwarding
  anomalies}.
\newblock {\em ACM SIGMETRICS Performance Evaluation Review}, 32(1):416--417,
  2004.

\bibitem{schulman2011pingin}
Aaron Schulman and Neil Spring.
\newblock {Pingin' in the rain}.
\newblock In {\em Proceedings of ACM Internet Measurement Conference}, pages
  19--28, 2011.

\bibitem{Tezzeract18}
M.~Syamkumar, S.~Mani, R.~Durairajan, P.~Barford, and J.~Sommers.
\newblock {Wrinkles in Time: Detecting Internet-wide Events via NTP}.
\newblock In {\em {Proceedings of the IFIP Networking}}, May 2018.

\bibitem{syamkumar16}
Meenakshi Syamkumar, Ramakrishnan Durairajan, and Paul Barford.
\newblock {Bigfoot: A Geo-based Visualization Methodology for Detecting BGP
  Threats}.
\newblock In {\em IEEE Symposium on Visualization for Cyber Security}, 2016.

\bibitem{turner2013comparison}
Daniel Turner, Kirill Levchenko, Stefan Savage, and Alex~C Snoeren.
\newblock {A comparison of syslog and IS-IS for network failure analysis}.
\newblock In {\em Proceedings of the ACM Internet Measurement Conference},
  pages 433--440, 2013.

\bibitem{turner2010california}
Daniel Turner, Kirill Levchenko, Alex~C Snoeren, and Stefan Savage.
\newblock {California fault lines: understanding the causes and impact of
  network failures}.
\newblock In {\em ACM SIGCOMM Computer Communication Review}, volume~40, pages
  315--326, 2010.

\bibitem{wang2006measurement}
Feng Wang, Zhuoqing~Morley Mao, Jia Wang, Lixin Gao, and Randy Bush.
\newblock A measurement study on the impact of routing events on end-to-end
  internet path performance.
\newblock In {\em ACM SIGCOMM Computer Communication Review}, volume~36, pages
  375--386. ACM, 2006.

\bibitem{Yu13}
M.~Yu, L.~Jose, and R.~Miao.
\newblock {Software Defined Traffic Measurement with OpenSketch}.
\newblock In {\em {Proceedings of the USENIX NSDI Conference}}, April 2013.

\bibitem{zhang2004planetseer}
Ming Zhang, Chi Zhang, Vivek~S Pai, Larry~L Peterson, and Randolph~Y Wang.
\newblock Planetseer: Internet path failure monitoring and characterization in
  wide-area services.
\newblock In {\em OSDI}, 2004.

\bibitem{zhu2012latlong}
Yaping Zhu, Benjamin Helsley, Jennifer Rexford, Aspi Siganporia, and Sridhar
  Srinivasan.
\newblock Latlong: Diagnosing wide-area latency changes for cdns.
\newblock {\em IEEE Transactions on Network and Service Management}, 2012.

\end{thebibliography}
\end{document}